\documentclass[review]{elsarticle}
\usepackage{amsmath,amssymb,amsthm}
\usepackage{mathtools}
\usepackage{graphicx}
\usepackage{algorithm}
\usepackage{algorithmic}
\usepackage{bbm}
\usepackage{bm}
\usepackage{multirow} 
\usepackage{booktabs} 
\usepackage{xcolor}
\newtheorem{theorem}{Theorem}

\newtheorem{definition}{Definition}

\newtheorem{remark}{Remark}
\usepackage[margin=1in]{geometry}
\begin{document}

\begin{frontmatter}

    \title{Universal Latent Homeomorphic Manifolds: \\
    A Framework for Cross-Domain Representation Unification}

    % Author 1
    \author[1]{Tong Wu\corref{cor1}}
    \ead{tong.wu@ucf.edu}
    
    % Author 2
    \author[1]{Tayab Uddin Wara}
     \ead{tayabuddin.wara@ucf.edu} % Optional

    % Author 3
    \author[1]{Daniel Hernandez}
     \ead{daniel.hernandez4@ucf.edu} % Optional
  
    % Author 4
    \author[1]{Sidong Lei}
    \ead{sidong.lei@ucf.edu}

    % Affiliations
    \address[1]{University of Central Florida, Orlando, USA}

    % Corresponding author marker
    \cortext[cor1]{Corresponding author.}

    \begin{abstract}
    We present the  {Universal Latent Homeomorphic Manifold} (ULHM), a framework that unifies semantic representations (e.g., human descriptions, diagnostic labels) and observation-driven machine representations (e.g., pixel intensities, sensor readings) into a single latent structure. Despite originating from fundamentally different pathways, both modalities capture the same underlying reality. We establish \emph{homeomorphism}, a continuous bijection preserving topological structure, as the mathematical criterion for determining when latent manifolds induced by different semantic-observation pairs can be rigorously unified. This criterion provides theoretical guarantees for three critical applications: (1)  {semantic-guided sparse recovery} from incomplete observations, (2)  {cross-domain transfer learning} with verified structural compatibility, and (3)  {zero-shot compositional learning} via valid transfer from semantic to observation space.
    Our framework learns continuous manifold-to-manifold transformations through conditional variational inference, avoiding brittle point-to-point mappings. We develop practical verification algorithms, including trust, continuity, and Wasserstein distance metrics, that empirically validate homeomorphic structure from finite samples. Experiments demonstrate: (1) sparse image recovery from 5\% of CelebA pixels and MNIST digit reconstruction at multiple sparsity levels, (2) cross-domain classifier transfer achieving 86.73\% accuracy from MNIST to Fashion-MNIST without retraining, and (3) zero-shot classification on unseen classes achieving 89.47\% on MNIST, 84.70\% on Fashion-MNIST, and 78.76\% on CIFAR-10. Critically, the homeomorphism criterion determines when different semantic-observation pairs share compatible latent structure, enabling principled unification into universal representations and providing a mathematical foundation for decomposing general foundation models into domain-specific components.
    \end{abstract}


\end{frontmatter}

\section{Introduction}
\subsection{Background and Motivation}
Deep learning models learn representations through pathways fundamentally different from semantic descriptions \cite{bengio2013representation, locatello2019challenging}. While traditional paradigms prioritize mapping data to specific task targets, they often lack the structural grounding inherent in semantic frameworks. Semantic information captures high-level conceptual structure through annotations. For example, describing a person as ``wearing glasses'' or ``having black hair,'' recognizing that handwritten digits 0, 6, 9 have one circle while 8 has two, or labeling medical images as ``irregular morphology.'' In contrast, deep learning systems extract hierarchical features directly from raw measurements (pixel intensities, time-series values, sensor outputs, or spectral data) through statistical pattern recognition. Despite these different origins, both representations aim to capture the same underlying reality \cite{huh2024position}. This divergence raises two fundamental questions: \emph{Can semantic and data-driven representations be unified into a single latent manifold where both converge to a shared geometric structure? Furthermore, under what mathematical conditions can we establish a universal latent manifold that enables a learning model to remain structurally consistent, providing a formal path for transitioning general-purpose foundation models to domain-specific experts?}

Answering these questions has direct implications for three practical representation learning challenges. While we primarily use image processing examples for illustration, the framework applies broadly to any domain where semantic information (such as human descriptions, state variables in control systems, or symbolic labels) coexists with raw measurements, including signal processing, network control, and sensor fusion. To address these diverse applications, our theoretical framework provides a rigorous foundation for overcoming the following three technical bottlenecks.

\begin{itemize}
\item \textbf{Sparse recovery with semantic priors.} During training, the model observes complete measurements paired with rich semantic descriptions. In contrast, deployment relies on highly undersampled measurements without accompanying semantic annotations. This creates an asymmetric scenario where high-level knowledge is available during learning but strictly absent during inference. This leads to a fundamental question: Can we establish a latent mapping that achieves effectively lossless compression, and what are the theoretical limits of such recovery? Specifically, can a learning model encode enough structural logic during training that a drastic reduction in data does not result in a loss of signal integrity, allowing the network to implicitly leverage high-level constraints to reconstruct full-dimensional signals from incomplete observations?

\item \textbf{Cross Domain Geometric Alignment and Domain Transfer.} Multiple related semantic observation pairs, such as clinical descriptions paired with either CT or MRI scans, intuitively share an underlying latent structure. However, no principled mathematical criterion exists to determine whether the manifolds induced by these heterogeneous pairs can be rigorously unified. Current ad hoc alignment methods remain fragile, often failing when distributions shift or new modalities emerge. This raises a fundamental question: Under what conditions can diverse semantic observation pairs be unified into a single geometric framework? Specifically, can we establish a universal latent structure that remains consistent across disparate domains, providing a formal basis for the cross modality transfer of knowledge?

\item \textbf{Zero-shot attribute composition.} Models learn semantic attributes from training examples where each attribute typically appears in isolation. However, training data rarely cover all $2^{K}$ possible combinations of $K$ attributes. At test time, images often exhibit novel combinations never jointly observed, creating a zero shot scenario where standard learning models fail because they treat attributes as independent entities without capturing their underlying structural relationships. This leads to a fundamental question: Can the latent manifold components be structured to enable valid zero shot composition within the observation space? Specifically, can the manifold capture the geometric logic of attribute interactions such that novel combinations remain consistent with the learned manifold structure, even when they were never explicitly seen during training?
\end{itemize}

We establish homeomorphism as the fundamental criterion for semantic observation unification. In our framework, each semantic observation pair induces a latent manifold through representation learning. When the latent manifolds originating from disparate pairs are homeomorphic, sharing an identical topological structure through a continuous and invertible bijection, they can be rigorously unified into a universal representation space. This condition provides a unified mathematical foundation for addressing the three previously identified challenges: semantic priors can provably guide sparse recovery, heterogeneous data can be integrated with formal guarantees for transfer learning, and compositional reasoning in the semantic space transfers validly to the observation space for zero shot learning.

\begin{figure*}[!htp]
  \vspace{-0.2cm}
  \centering
    % Replace 'ulhm_workflow.pdf' with your actual filename
    \includegraphics[width=0.95\textwidth]{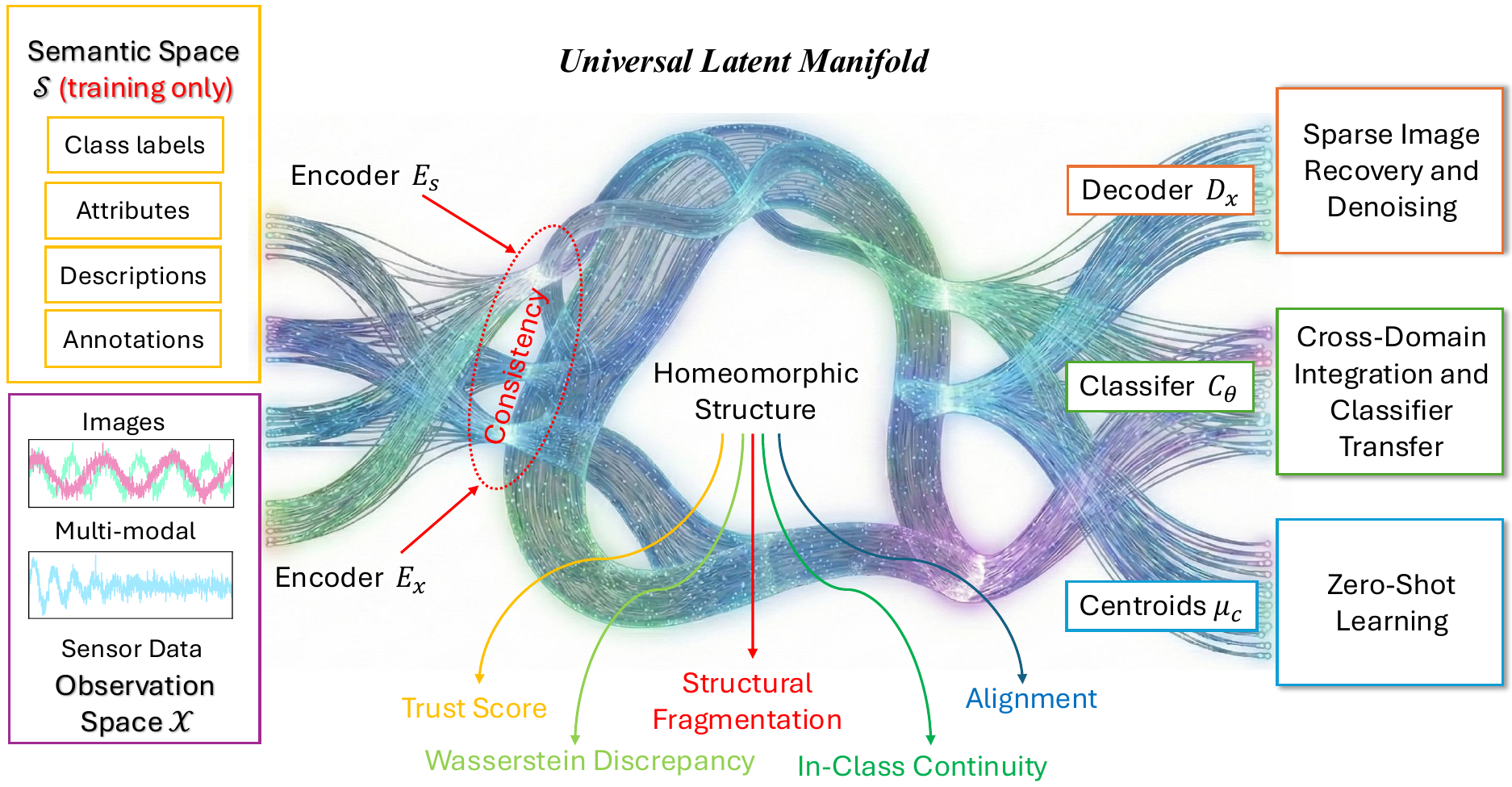}
    \vspace{-0.5cm}
  \caption{Overview of the proposed Universal Latent Manifold (ULHM) framework. 
  The architecture bridges the \textbf{Semantic Space} $\mathcal{S}$ (utilized during training with class labels, attributes, and descriptions) and the \textbf{Observation Space} $\mathcal{X}$ (comprising images, multi-modal inputs, and sensor data). 
  This integration enables diverse downstream capabilities, including Sparse Image Recovery and Denoising, Cross-Domain Integration and Transfer, and Zero-Shot Learning.}
  \label{fig:ulhm_workflow}
  \vspace{-0.5cm}
\end{figure*}

\subsection{Related Work}
We review prior work on latent geometry and manifold-based embeddings, organizing the discussion around five key perspectives that align with our methodological contributions: universal geometry of embeddings, sparse recovery with semantic priors, cross-domain transfer learning, and zero-shot compositional generalization.

\subsubsection{Universal Latent Manifold} Existing approaches to universal representations remain incomplete. Recent empirical work on the  Platonic Representation Hypothesis~\cite{huh2024position} observes that neural networks trained on different modalities converge toward aligned geometric structures. While subsequent research has provided information-geometric perspectives~\cite{lobashev2025information}, evidence of universal weight subspaces~\cite{kaushik2025universal}, and methods to harness universal embedding geometries~\cite{jha2025harnessing}, these observations lack a rigorous formalization of \emph{when} such alignment occurs and \emph{why} it succeeds or fails. Alignment methods based on optimal transport~\cite{alvarez2018gromov,grave2019unsupervised,chen2020graph} and representation similarity metrics~\cite{kornblith2019similarity} can match embedding spaces across modalities, yet they provide no criterion for determining structural compatibility prior to alignment. Similarly, multi-view learning methods~\cite{wang2015deep} construct shared representations but assume fusion is always valid, which leads to failure when underlying structures are fundamentally incompatible. Furthermore, while recent advances in manifold learning with topological and geometric regularization~\cite{wang2025latent,rhodes2025guided} and analyses of neural representation topology~\cite{lin2024topology} have improved our understanding of latent structure, these approaches typically focus on reconstruction and local geometry without the explicit semantic integration required for compositional reasoning.

\textbf{Our Difference:} The critical gap in these existing methods is that they analyze input and output distributions separately. For instance, two semantic-observation pairs may possess identical input geometries but map to opposite output labels (e.g., high intensity representing ``positive" in one modality and ``negative" in another). Even if both form manifolds of the same shape, existing approaches cannot distinguish whether they can be meaningfully unified. Our framework addresses this by analyzing the \emph{joint} structure of semantic-observation pairs. We establish \emph{homeomorphism} not just as a property to be observed, but as the rigorous mathematical criterion to determine \emph{validity}, ensuring that unification is performed only when the latent manifolds induced by different pairs are topologically equivalent.

\subsubsection{Homeomorphic Learning}
The concept of homeomorphism has been increasingly utilized to model structural equivalence in machine learning. In unsupervised domain adaptation, Zhou et al.~\cite{zhou2023homeomorphism} employ invertible neural networks to construct a homeomorphic alignment between source and target feature distributions, ensuring that discriminative structures are preserved during transfer. In the context of optimization, homeomorphic projection methods~\cite{liang2024homeomorphic} map unconstrained neural network outputs onto manifolds defined by physical requirements to ensure solution validity. Other approaches leverage homeomorphism to preserve intrinsic data topology during dimensionality reduction~\cite{moor2020topological} or to resolve the manifold mismatch between complex data distributions and simple latent priors in variational autoencoders~\cite{falorsi2018explorations}. Earlier work on Homeomorphic Manifold Analysis~\cite{elgammal2013homeomorphic} utilized these mappings to separate style and content within a single domain. However, these methods typically treat homeomorphism as a mapping tool to align marginal distributions or enforce \textbf{assumed latent topologies}. By relying on these predefined constraints, such approaches effectively force the data to fit a geometric shape guessed beforehand (e.g., a unit circle or torus), rather than discovering the true unified structure inherent in the joint semantic-observation relationship. For instance, two datasets may have identical feature distributions but opposite label semantics (e.g., high temperature indicating `stable' in one system but `critical' in another).  Our framework addresses this by establishing homeomorphism not merely as a mapping mechanism, but as a rigorous criterion for the \emph{joint semantic-observation structure}, enabling us to reject unification when the underlying mappings are topologically incompatible despite geometric similarity.

\subsubsection{Sparse Recovery and Compressed Sensing}
Traditional compressed sensing methods recover sparse signals using hand-crafted priors such as wavelet bases or total variation. While foundational deep generative approaches~\cite{bora2017compressed} and unsupervised methods like Deep Image Prior~\cite{ulyanov2018deep} demonstrated that learnable priors could effectively model reconstruction mappings, recent advancements have shifted toward score-based generative models~\cite{song2022solving} and diffusion restoration methods~\cite{kawar2022denoising, chung2023diffusion}, which solve inverse problems by iteratively refining noisy distributions, and transformer-based architectures~\cite{liang2021swinir} that exploit long-range dependencies for restoration. Although some recent works incorporate semantic information  (such as semantic-aware sensing matrices~\cite{zhang2023semantic} or latent diffusion models conditioned on explicit semantic inputs~\cite{rombach2022high}), these approaches typically lack rigorous topological guarantees and often require semantic availability during inference. In contrast, our approach verifies that the mapping between observation and latent spaces forms a homeomorphism, ensuring lossless topological compression. This homeomorphic structure allows semantic priors learned during training to implicitly guide sparse reconstruction at test time without explicit semantic input.

% \subsubsection{Universal Manifold Classification and Transfer}
% Recent advances in transfer learning have shifted from full fine-tuning to parameter-efficient modularity. Techniques such as Low-Rank Adaptation (LoRA)~\cite{hu2021lora} and adapter-based methods~\cite{houlsby2019parameter} enable classification transfer by optimizing sparse updates within a fixed weight space. Concurrently, model merging approaches like Task Arithmetic~\cite{ilharco2023editing} and TIES-Merging~\cite{yadav2023ties} demonstrate that classification capabilities can be transferred or combined by linearly manipulating model parameters, a phenomenon recently attributed to the emergence of Universal Weight Subspaces~\cite{kaushik2025universal}. While these methods achieve remarkable efficiency—often reducing trainable parameters by orders of magnitude—they operate under the assumption that the source and target tasks share a compatible underlying geometry~\cite{jha2025harnessing}. However, they lack a mechanism to detect topological mismatches: if the source and target manifolds differ in connectivity or intrinsic dimensionality, weight-space interpolation yields invalid decision boundaries. Our framework provides the missing topological guarantee. By verifying that the latent manifolds are homeomorphic, we ensure that the classifier's decision boundaries can be continuously deformed from the source to the target domain, enabling rigorous zero-shot transfer and fusion with theoretical stability.

\subsubsection{Cross-Domain Transfer and Universal Classifiers}
Recent advances in transfer learning have shifted from full fine-tuning toward parameter-efficient modularity. Techniques such as Low-Rank Adaptation (LoRA)\cite{hu2021lora} and adapter-based methods\cite{houlsby2019parameter} enable domain transfer by optimizing sparse updates within fixed weight spaces. Concurrently, model merging approaches like Task Arithmetic~\cite{ilharco2023editing} and TIES-Merging~\cite{yadav2023ties} demonstrate that learned capabilities can be transferred or combined through linear parameter manipulation. This transferability has been attributed to the emergence of universal weight subspaces~\cite{kaushik2025universal}, with extensions to graph learning frameworks that accommodate topological variations~\cite{wu2025universal}.

While these methods achieve remarkable parameter efficiency, often reducing trainable parameters by orders of magnitude, they implicitly assume that source and target domains share compatible underlying geometry~\cite{jha2025harnessing}. Critically, even methods designed to handle structural variations lack explicit mechanisms to \emph{verify} topological compatibility before attempting transfer. Our framework addresses this gap by establishing homeomorphism verification as a prerequisite for valid transfer. When latent manifolds are verified to be homeomorphic, classifier decision boundaries can be continuously deformed across domains, enabling principled zero-shot transfer with theoretical guarantees rather than empirical heuristics.

\subsubsection{Zero-Shot Learning and Compositional Generalization}
Zero-shot learning methods recognize unseen classes by composing semantic attributes, a paradigm established by foundational attribute-based classification models~\cite{pourpanah2022review, lampert2013attribute}. To overcome the bias toward seen classes inherent in these mapping-based approaches, generative frameworks~\cite{xian2018feature} were developed to synthesize visual features for unseen categories, effectively converting zero-shot tasks into supervised learning problems. However, standard generative models often struggle with \emph{compositional generalization}, failing to capture the causal structure required to distinguish valid attribute-object combinations from spurious correlations~\cite{atzmon2020causal}. Recent state-of-the-art methods have addressed this limitation through compositional soft prompting~\cite{nayak2023learning}, which optimizes learnable context vectors to disentangle attributes in vision-language models, and through diffusion-based classifiers~\cite{li2023your} that exploit density estimation for robust zero-shot recognition. Despite these advancements, these approaches rely on statistical alignment without topological guarantees. Our framework addresses this by verifying homeomorphic structure in the latent manifold, ensuring that unseen classes occupy geometrically separated regions through preserved topology, enabling zero-shot classification via nearest-centroid assignment without requiring explicit supervision on unseen classes.

\subsection{Contributions}
To address these challenges, we propose the manifold learning framework illustrated in Fig.~\ref{fig:ulhm_workflow}, which unifies semantic descriptions and raw observations through verified homeomorphic structure.  The key insight is that when semantic information and observations originate from the same underlying reality, their latent representations should be homeomorphic, preserving topological structure through continuous bijection. We establish homeomorphism as the rigorous mathematical criterion for determining when latent manifolds from different semantic-observation pairs can be validly unified. Unlike existing methods that analyze distributions separately without principled verification, our framework provides practical algorithms to empirically validate homeomorphic structure from finite samples, enabling us to reject incompatible unifications before deployment. The main contributions of this work are as follows.

\begin{itemize}
\item \textbf{Asymmetric semantic-observation manifold learning.} We propose a framework that learns from semantic-observation pairs during training but performs inference from observations alone at test time. By embedding semantic structure as geometric constraints in the latent manifold through joint reconstruction and consistency objectives, the framework enables semantic priors to implicitly guide reconstruction and reasoning without requiring semantic annotations at deployment. This asymmetric paradigm, which leverages semantic supervision during training while maintaining observation-only inference, provides a principled approach to incorporating high-level knowledge into observation-driven systems.

\item \textbf{Homeomorphism as the criterion for valid semantic-observation unification.} We establish homeomorphism as the mathematical condition under which latent manifolds induced by different semantic-observation pairs can be rigorously unified. This criterion ensures that the joint relational structure between semantics and observations is preserved through continuous bijection, providing theoretical guarantees for when cross-domain transfer, sparse recovery with semantic priors, and compositional reasoning remain valid.

\item \textbf{Sparse recovery via homeomorphic compression and theoretical limits.} We establish that when the encoder-decoder mapping approximates a homeomorphism, the latent representation serves as a lossless topological compression preserving the structural integrity of signals. This homeomorphic property enables us to identify the fundamental limits of compression: the framework can reconstruct full-dimensional signals from highly undersampled observations by traversing the manifold's intrinsic geometry. Verified homeomorphism thus provides both a theoretical criterion for compression limits and practical superior performance in sparse recovery without requiring semantic annotations at test time.

\item \textbf{Verified homeomorphism enables zero-shot cross-domain transfer.} We establish that verified homeomorphism provides theoretical guarantees for valid cross-domain integration and classifier transfer. A universal classifier trained on one domain transfers directly to heterogeneous domains without retraining, substantially outperforming existing domain adaptation methods and validating that homeomorphic verification correctly predicts when zero-shot transfer will succeed.

\item \textbf{Zero-shot classification via topological preservation.} We demonstrate that homeomorphic structure enables zero-shot recognition of unseen classes through nearest-centroid classification in the learned latent manifold. By preserving topological structure, the framework achieves superior performance compared to state-of-the-art zero-shot learning methods across multiple datasets, validating that preserved topology enables geometric separation without explicit supervision on unseen classes.

\end{itemize}

The remainder of this paper is organized as follows. Section II presents the mathematical framework for universal latent homeomorphic manifolds. Section III develops the manifold learning architecture and homeomorphism verification methods. Section IV presents experimental results demonstrating the framework's effectiveness on sparse recovery, heterogeneous data integration, and zero-shot learning. Section V concludes with discussions and future directions.

\section{Universal Latent Homeomorphic Manifold Framework}
We present the mathematical framework for learning universal latent homeomorphic manifolds that unify semantic and observation representations. The key insight is treating each semantic-observation pair as inducing a latent manifold through representation learning, then verifying whether manifolds from different pairs are homeomorphic. When homeomorphism holds, the manifolds share topological structure and can be unified into a universal representation space.

\subsection{Problem Formulation and Notation}
Consider a scenario where we have access to heterogeneous information about the same underlying phenomenon: \emph{semantic descriptions} and \emph{raw observations}. These information sources can take multiple forms depending on the application domain. For handwritten digit recognition, semantic descriptions might include structural attributes like ``has one closed loop'' alongside class labels, while raw observations consist of pixel intensities from MNIST images. For medical imaging reconstruction, observations might include sparse measurements and sampling masks, while semantics encompass ground truth complete images and diagnostic annotations. For multi-sensor systems, observations may involve heterogeneous sensor readings at different spatial or temporal resolutions, while semantics provide state labels or physical parameters.

We formally define the following notation to accommodate this heterogeneity. Let $\mathcal{X} = \{\mathcal{X}^{(1)}, \mathcal{X}^{(2)}, \ldots, \mathcal{X}^{(M)}\}$ denote a collection of observation spaces, where each $\mathcal{X}^{(m)}$ represents a distinct type of measurement. For example:
\begin{itemize}
    \item $\mathcal{X}^{(1)} \subset \mathbb{R}^{D_1}$: sparse or incomplete measurements (e.g., sampled pixel values, sensor readings)
    \item $\mathcal{X}^{(2)} \subset \{0,1\}^{D_2}$: observation mask matrix indicating which measurements are available
    \item $\mathcal{X}^{(3)} \subset \mathbb{R}^{D_3}$: auxiliary measurements from complementary modalities
\end{itemize}
We denote $\mathbf{x} = \{x^{(1)}, x^{(2)}, \ldots, x^{(M)}\}$ as a multi-modal observation tuple, where $x^{(m)} \in \mathcal{X}^{(m)}$. Not all modalities need be present for every sample.
Similarly, let $\mathcal{S} = \{\mathcal{S}^{(1)}, \mathcal{S}^{(2)}, \ldots, \mathcal{S}^{(L)}\}$ denote a collection of semantic spaces capturing different forms of high-level information. For example:
\begin{itemize}
    \item $\mathcal{S}^{(1)} \subset \mathbb{R}^{H \times W \times C}$: ground truth complete signals or images
    \item $\mathcal{S}^{(2)} \subset \{0,1\}^K$: binary attribute vectors or multi-label annotations
    \item $\mathcal{S}^{(3)} \subset \mathbb{R}^{D_{\text{text}}}$: text embeddings from natural language descriptions
    \item $\mathcal{S}^{(4)} \subset \mathbb{Z}_+$: discrete class labels or categorical variables
\end{itemize}
We denote $\mathbf{s} = \{s^{(1)}, s^{(2)}, \ldots, s^{(L)}\}$ as a multi-modal semantic tuple, where $s^{(l)} \in \mathcal{S}^{(l)}$.

Let $\mathcal{Z} \subset \mathbb{R}^d$ denote the universal latent space containing the \emph{latent manifold} $\mathcal{M}_z \subset \mathcal{Z}$, where $z \in \mathcal{M}_z$ represents learned representations.  We denote $(\mathbf{x}_i, \mathbf{s}_i)$ as a paired sample, where multi-modal observations $\mathbf{x}_i$ are associated with multi-modal semantic information $\mathbf{s}_i$, and $\mathcal{D} = \{(\mathbf{x}_i, \mathbf{s}_i)\}_{i=1}^N$ as the training dataset of $N$ paired samples. When considering multiple datasets from different sources, we use bracket notation $\mathcal{D}^{[k]} = \{(\mathbf{x}_i^{[k]}, \mathbf{s}_i^{[k]})\}_{i=1}^{N_k}$ to distinguish dataset $k$, where the bracket superscript $[\cdot]$ denotes dataset index and the parenthesis superscript $(\cdot)$ denotes modality index within a dataset.

\paragraph{The Role of the Latent Manifold} While high-dimensional data may concentrate near low-dimensional geometric structures under the manifold hypothesis, the learned latent manifold $\mathcal{M}_z \subset \mathcal{Z}$ serves a fundamentally different purpose than mere dimensionality reduction. The latent space is designed to be a \emph{standardized and informationally enriched} representation that integrates structure from both observation space $\mathcal{X}$ and semantic space $\mathcal{S}$. During training, the framework learns to embed paired semantic-observation information $(\mathbf{s}, \mathbf{x})$ into $\mathcal{M}_z$, creating a representation that is simultaneously:
\begin{itemize}
    \item \emph{Standardized}: Different modalities and datasets map to a common geometric structure
    \item \emph{Semantically enriched}: Captures high-level conceptual relationships encoded in $\mathbf{s}$
    \item \emph{Observation-grounded}: Preserves discriminative information from raw measurements $\mathbf{x}$
\end{itemize}

Critically, at test time, the latent representation can be inferred from observations alone via encoder mappings $E_x^{(m)}: \mathcal{X}^{(m)} \to \mathcal{M}_z$, even when semantic information is unavailable. The semantic structure learned during training becomes implicitly embedded in the latent geometry, enabling the observation encoders to map to semantically meaningful regions of $\mathcal{M}_z$ without requiring explicit semantic input. This asymmetry, which uses both modalities during training but only observations during testing, is key to the framework's practical utility.

\paragraph{The Universal Manifold Problem} Given multiple datasets with heterogeneous domains, for instance, $\mathcal{D}^{[1]}$ containing semantic attributes paired with MNIST handwritten digits and $\mathcal{D}^{[2]}$ containing semantics paired with Fashion-MNIST clothing items, the fundamental question is: \emph{under what conditions can these heterogeneous semantic-observation pairs be unified into a single universal latent manifold $\mathcal{M}_z$ that preserves the joint relational structure of each dataset?}

The challenge is threefold. \textbf{First}, we must learn encoder mappings $\{E_x^{(m)}, E_s^{(l)}\}$ that embed both semantics and observations from different domains into a shared latent manifold $\mathcal{M}_z$ while preserving their joint topological structure. \textbf{Second}, we must learn decoder mappings $\{D_\theta^{(m)}, D_\theta^{(l)}\}$ parameterized by $\theta$ that faithfully reconstruct each modality from latent codes, ensuring the latent representation captures the semantic-observation correspondence. \textbf{Third}, and most critically, we must establish a mathematical criterion to determine \emph{when such unification is valid}, that is, when the latent manifolds induced by different semantic-observation pairs are topologically compatible for fusion.

Our key insight is that successful unification requires these latent manifolds to be \textbf{homeomorphic}, preserving the relational structure between semantics and observations through continuous bijection. This homeomorphic criterion provides theoretical guarantees for three critical capabilities: semantic-guided sparse recovery from incomplete observations, zero-shot cross-domain classifier transfer, and compositional zero-shot learning. We formalize this homeomorphism criterion in Subsection~\ref{sec:homeomorphism} and develop practical verification methods in Algorithm~\ref{alg:homeomorphism}.

\paragraph{Notation Summary}
For clarity, we summarize the key notation used throughout:
\begin{itemize}
    \item \textbf{Superscripts}: $(m)$ = observation modality index, $(l)$ = semantic modality index, $[k]$ = dataset index
    \item \textbf{Subscripts}: $i, j$ = sample indices, $\theta, \phi$ = neural network parameters
    \item \textbf{Spaces}: $\mathcal{X}^{(m)}$ = observation space for modality $m$, $\mathcal{S}^{(l)}$ = semantic space for modality $l$, $\mathcal{Z}$ = universal latent space (ambient), $\mathcal{M}_z \subset \mathcal{Z}$ = latent manifold
    \item \textbf{Mappings}: $E_x^{(m)}: \mathcal{X}^{(m)} \to \mathcal{M}_z$ = observation encoder, $E_s^{(l)}: \mathcal{S}^{(l)} \to \mathcal{M}_z$ = semantic encoder, $D_\theta^{(m)}: \mathcal{M}_z \to \mathcal{X}^{(m)}$ = observation decoder
\end{itemize}

\subsection{Latent Manifold Learning Architecture}
We learn latent manifold representations that explicitly capture how semantic-observation correspondences induce geometric structure in latent space. The framework maps high-dimensional observations and semantic information onto a continuous latent manifold $\mathcal{M}_z$ where geometric proximity simultaneously reflects data similarity and semantic relatedness.

\subsubsection{Latent Manifold Representation}
Our goal is to learn a latent manifold $\mathcal{M}_z \subset \mathbb{R}^d$ that serves as a common geometric space where both observation modalities from $\mathcal{X}$ and semantic modalities from $\mathcal{S}$ can be embedded while preserving their intrinsic structure. We learn encoder mappings that project data onto this latent manifold:
\begin{equation}
E_x^{(m)}: \mathcal{X}^{(m)} \to \mathcal{M}_z, \quad E_s^{(l)}: \mathcal{S}^{(l)} \to \mathcal{M}_z
\end{equation}
implemented as neural networks parameterized by $\phi$. For a given input, the latent representation is obtained deterministically as $z = E_x^{(m)}(x^{(m)}; \phi)$ or $z = E_s^{(l)}(s^{(l)}; \phi)$.

Given a latent code $z \in \mathcal{M}_z$, we define conditional likelihood models for reconstructing each modality. The reconstruction model for observations is
\begin{equation}
p_{\theta}(x^{(m)}|z) = \mathcal{N}(D_{\theta}^{(m)}(z), \sigma_{x}^2 I)
\end{equation}
where $D_{\theta}^{(m)}: \mathcal{M}_z \to \mathcal{X}^{(m)}$ is a neural network decoder parameterized by $\theta$. For semantic modalities, we use appropriate likelihood functions:
\begin{equation}
p_{\theta}(s^{(l)}|z) = \begin{cases}
 \prod_{j} \text{Bernoulli}(D_{\theta}^{(l)}(z)_j)  & \text{binary} \\
\mathcal{N}(D_{\theta}^{(l)}(z), \sigma_{s}^2 I) & \text{continuous} \\
\text{Categorical}(D_{\theta}^{(l)}(z)) & \text{categorical} \\
\prod_{t=1}^{T} p(s_t^{(l)}|z, s_{<t}^{(l)}) & \text{sequential}
\end{cases}
\end{equation}
where the sequential case handles variable-length tokenized representations (e.g., annotation sequences, text descriptions) through autoregressive modeling. All encoders and decoders are implemented as Lipschitz continuous neural networks to ensure stable manifold mappings.

\subsubsection{Encoder and Decoder Architecture}

We learn encoder mappings that project observations and semantics onto the shared latent manifold $\mathcal{M}_z$, and decoder mappings that reconstruct each modality from latent codes.

\paragraph{Encoder Networks} For each observation modality $m$, we define the encoder as a neural network:
\begin{equation}
z = E_x^{(m)}(x^{(m)}; \phi): \mathcal{X}^{(m)} \to \mathcal{M}_z \subset \mathbb{R}^d
\end{equation}
parameterized by $\phi$, which deterministically maps observations to latent codes. Similarly, for semantic modalities $l$, we define encoders:
\begin{equation}
z = E_s^{(l)}(s^{(l)}; \phi): \mathcal{S}^{(l)} \to \mathcal{M}_z \subset \mathbb{R}^d
\end{equation}

\paragraph{Decoder Networks} For each modality, we define decoders that reconstruct from latent codes. For observation modality $m$:
\begin{equation}
\hat{x}^{(m)} = D_{\theta}^{(m)}(z): \mathcal{M}_z \to \mathcal{X}^{(m)}
\end{equation}
parameterized by $\theta$. Similarly, for semantic modalities $l$:
\begin{equation}
\hat{s}^{(l)} = D_{\theta}^{(l)}(z): \mathcal{M}_z \to \mathcal{S}^{(l)}
\end{equation}

When a sample contains multiple modalities, we denote the available observation modalities as $\mathcal{M}_i^x \subseteq \{1, \ldots, M\}$ and available semantic modalities as $\mathcal{M}_i^s \subseteq \{1, \ldots, L\}$ for sample $i$. The framework naturally handles missing modalities: if modality $m \notin \mathcal{M}_i^x$, we simply do not compute $E_x^{(m)}(x_i^{(m)})$ for that sample.

This manifold-to-manifold perspective differs fundamentally from point-to-point mappings: the encoders $E_x^{(m)}$ and $E_s^{(l)}$ are continuous and preserve neighborhood relationships, meaning nearby points on the data manifolds map to nearby points on the latent manifold. This topological preservation property is enforced through the local consistency loss.

\subsubsection{Multi-Objective Training}
We train the framework through a composite objective that simultaneously ensures reconstruction fidelity, cross-modal alignment, and topological preservation. For a dataset $\mathcal{D} = \{(\mathbf{x}_i, \mathbf{s}_i)\}_{i=1}^N$, the complete loss is
\begin{equation}
\mathcal{L}_{\text{ULHM}} = \mathcal{L}_{\text{recon}}^x + \mathcal{L}_{\text{recon}}^s + \lambda_c \mathcal{L}_{\text{consist}} + \lambda_\ell \mathcal{L}_{\text{local}} + \lambda_p \mathcal{L}_{\text{percep}}
\label{eq:ulhm_loss}
\end{equation}
where $\lambda_c, \lambda_\ell, \lambda_p \geq 0$ are hyper-parameters. We now detail each component.

\paragraph{Reconstruction Objectives} For each modality with available data, we minimize the negative log-likelihood of the reconstruction. The observation reconstruction loss is
\begin{equation}
\mathcal{L}_{\text{recon}}^x = \sum_{i=1}^N \sum_{m \in \mathcal{M}_i^x} -\log p_{\theta}(x_i^{(m)}|z_i^{(m)})
\end{equation}
where $z_i^{(m)} = E_x^{(m)}(x_i^{(m)}; \phi)$ is the latent encoding. Similarly, the semantic reconstruction loss is
\begin{equation}
\mathcal{L}_{\text{recon}}^s = \sum_{i=1}^N \sum_{l \in \mathcal{M}_i^s} -\log p_{\theta}(s_i^{(l)}|z_i^{(l)})
\end{equation}
where $z_i^{(l)} = E_s^{(l)}(s_i^{(l)}; \phi)$. For Gaussian likelihoods, $-\log p_{\theta}(x|z) \propto \|D_{\theta}^{(m)}(z) - x\|_2^2$, which corresponds to mean squared error. The reconstruction terms ensure that points on the latent manifold decode back to realistic samples on the data manifolds.

\paragraph{Cross-Modal Consistency} To learn a shared latent manifold rather than separate manifolds for each modality, we enforce that different modalities from the same sample map to nearby points through a composite consistency objective combining both angular and magnitude alignment:
\begin{align}
\mathcal{L}_{\text{consist}} = &\lambda_{\text{cos}} \sum_{i=1}^N \Bigg[ \sum_{\substack{m, m' \in \mathcal{M}_i^x \\ m < m'}} \left(1 - \frac{z_i^{(m)} \cdot z_i^{(m')}}{\|z_i^{(m)}\|_2 \|z_i^{(m')}\|_2}\right) + \sum_{\substack{m \in \mathcal{M}_i^x \\ l \in \mathcal{M}_i^s}} \left(1 - \frac{z_i^{(m)} \cdot z_i^{(l)}}{\|z_i^{(m)}\|_2 \|z_i^{(l)}\|_2}\right) \Bigg] \nonumber \\
&+ \lambda_{\text{eucl}} \sum_{i=1}^N \Bigg[ \sum_{\substack{m, m' \in \mathcal{M}_i^x \\ m < m'}} \|z_i^{(m)} - z_i^{(m')}\|_2^2 + \sum_{\substack{m \in \mathcal{M}_i^x \\ l \in \mathcal{M}_i^s}} \|z_i^{(m)} - z_i^{(l)}\|_2^2 \Bigg]
\end{align}
where $z_i^{(m)} = E_x^{(m)}(x_i^{(m)}; \phi)$ and $z_i^{(l)} = E_s^{(l)}(s_i^{(l)}; \phi)$, and $\lambda_{\text{cos}}, \lambda_{\text{eucl}} \geq 0$ control the strength of angular and magnitude-based alignment, respectively. The cosine similarity term (weighted by $\lambda_{\text{cos}}$) enables directional alignment while permitting magnitude differences, which is essential for reconstruction tasks with sparse observations where information density varies significantly between modalities. The Euclidean distance term (weighted by $\lambda_{\text{eucl}}$) provides additional geometric constraints on embedding magnitudes, ensuring tighter clustering in scenarios where magnitude differences carry semantic meaning. For homogeneous multi-modal scenarios with comparable information density (e.g., multi-view imaging, synchronized sensor fusion, or fully-paired text-image data), we set both $\lambda_{\text{cos}}, \lambda_{\text{eucl}} > 0$ to enforce both angular and magnitude alignment. For heterogeneous scenarios with asymmetric information content (e.g., sparse vs. complete observations), we set $\lambda_{\text{eucl}} = 0$ and $\lambda_{\text{cos}} > 0$ to prioritize directional consistency. Geometrically, this composite loss constrains embeddings from all available modalities to overlap significantly in the latent space, creating a unified latent manifold $\mathcal{M}_z$.

\paragraph{Topological Preservation} Preserving the topological structure of the data manifolds is critical for learning meaningful representations. We enforce that local neighborhood structure is preserved through
\begin{align}
\mathcal{L}_{\text{local}} = &\sum_{m=1}^M \sum_{i: m \in \mathcal{M}_i^x} \sum_{j \in \mathcal{N}_\kappa(i, m)} \|z_i^{(m)} - z_j^{(m)}\|_2^2 +\sum_{l=1}^L \sum_{i: l \in \mathcal{M}_i^s} \sum_{j \in \mathcal{N}_\kappa(i, l)} \|z_i^{(l)} - z_j^{(l)}\|_2^2
\end{align}
where $\mathcal{N}_\kappa(i, m)$ denotes the indices of the $\kappa$-nearest neighbors of sample $i$ in modality $m$'s data space, and $z_i^{(m)} = E_x^{(m)}(x_i^{(m)}; \phi)$, $z_j^{(m)} = E_x^{(m)}(x_j^{(m)}; \phi)$. This loss acts as a manifold regularizer: if two samples are connected by a short geodesic path on the data manifold, they should also be nearby on the latent manifold.

\paragraph{Perceptual Quality} For observation modalities in the image domain, we employ a perceptual loss:
\begin{equation}
\mathcal{L}_{\text{percep}} = \sum_{m \in \mathcal{I}_{\text{img}}} \sum_{i: m \in \mathcal{M}_i^x} \sum_{r=1}^{L_{\text{feat}}} w_r \|\Phi_r(\hat{x}_i^{(m)}) - \Phi_r(x_i^{(m)})\|_2^2
\end{equation}
where $\hat{x}_i^{(m)} = D_{\theta}^{(m)}(z_i^{(m)})$ with $z_i^{(m)} = E_x^{(m)}(x_i^{(m)}; \phi)$, $\mathcal{I}_{\text{img}} \subseteq \{1,\ldots,M\}$ denotes the subset of image modality indices, $\Phi_r(\cdot)$ extracts features from layer $r$ of a pretrained VGG network, $L_{\text{feat}}$ is the number of feature layers, and $w_r > 0$ are layer-specific weights.

\subsection{Homeomorphism Verification Methods}\label{sec:homeomorphism}
We establish the conditions under which latent manifolds from different datasets can be rigorously unified. For datasets $\mathcal{D}^{[1]}$ and $\mathcal{D}^{[2]}$ (e.g., MNIST and Fashion-MNIST with shared attributes), both map to latent representations, but when can they share the same latent space while preserving structure? We prove that homeomorphism is the necessary and sufficient criterion for valid unification.

\subsubsection{Theoretical Analysis}

We establish strict conditions under which disjoint datasets form a unified, topologically consistent latent manifold. Central to our framework is the requirement that the representation learning process maintains a homeomorphism between the data space and the latent space, ensuring that no topological information is lost during compression.

\begin{definition}[Bi-Lipschitz Map]\label{def:bilipschitz}
A map $f: (\mathcal{M}_x, d_X) \to (\mathcal{M}_z, d_Z)$ is $(c_1, c_2)$-bi-Lipschitz if there exist constants $0 < c_1 \leq c_2 < \infty$ such that $\forall x, y \in \mathcal{M}_x$:
\begin{equation}
    c_1 d_X(x, y) \leq d_Z(f(x), f(y)) \leq c_2 d_X(x, y)
\end{equation}
where $d_X: \mathcal{M}_x \times \mathcal{M}_x \to \mathbb{R}_{\geq 0}$ and $d_Z: \mathcal{M}_z \times \mathcal{M}_z \to \mathbb{R}_{\geq 0}$ denote the geodesic distance metrics on the respective manifolds.
\end{definition}

\begin{theorem}\label{thm:unification}
\textbf{(Topological Unification via Latent Identifications).}
\textit{
Let $\{\mathcal{M}_x^{[k]}\}_{k=1}^K$ be pairwise disjoint topological spaces with continuous encoders
$E^{[k]}:\mathcal{M}_x^{[k]}\to \mathbb{R}^d$. Define the latent images
$\mathcal{M}_z^{[k]} := E^{[k]}(\mathcal{M}_x^{[k]})$ and unified latent support
$\mathcal{U}:=\bigcup_{k=1}^K \mathcal{M}_z^{[k]} \subset \mathbb{R}^d$.
Let $X:=\bigsqcup_{k=1}^K \mathcal{M}_x^{[k]}$ (disjoint union, coproduct topology) and define
$E:X\to \mathcal{U}$ by $E|_{\mathcal{M}_x^{[k]}}=E^{[k]}$.
Define $x\sim x'$ iff $E(x)=E(x')$.
Then the induced map $\widetilde{E}:X/\!\sim\ \to\ \mathcal{U}$ is a homeomorphism provided:
\begin{enumerate}
    \item[(i)] \emph{(Local Homeomorphism)} Each $E^{[k]}$ is a homeomorphism onto $\mathcal{M}_z^{[k]}$ (subspace topology from $\mathbb{R}^d$).
    \item[(ii)] \emph{(Coherence)} For all $V\subseteq \mathcal{U}$: $V$ is open in $\mathcal{U}$ iff $(E^{[k]})^{-1}(V)$ is open in $\mathcal{M}_x^{[k]}$ for all $k$.
\end{enumerate}
}
\end{theorem}

\begin{theorem}[Bi-Lipschitz Sufficiency]\label{thm:sufficiency}
Let $(\mathcal{M}_x, d_X)$ and $(\mathcal{M}_z, d_Z)$ be metric spaces with induced topologies. 
If encoder $E: \mathcal{M}_x \to \mathcal{M}_z$ is $(c_1, c_2)$-bi-Lipschitz with $c_1 > 0$, 
then $E$ is a homeomorphism onto its image, satisfying Condition (i) of Theorem \ref{thm:unification}.
\end{theorem}
\subsubsection{Empirical Verification Protocol}

Direct computation of the bi-Lipschitz constants $c_1, c_2$ and verification of the open cover condition (Theorem~\ref{thm:unification}, Condition ii) is intractable for real datasets. We bridge theory and practice through a hierarchical verification framework that employs computable metrics to detect violations of the theoretical conditions across structural, geometric, and homeomorphic dimensions.

\medskip
\noindent\textbf{Verification Strategy.}
While Theorems~\ref{thm:unification} and~\ref{thm:sufficiency} provide guarantees under ideal conditions (continuous maps, infinite samples), we verify their empirical satisfaction through complementary metrics organized in a hierarchical dependency structure. Each metric targets a specific failure mode:
\begin{enumerate} 
    \item \textbf{Structural Fragmentation:} $\beta_0(\mathcal{Z}_{\text{total}}) > 1$ with high pairwise $W_2$ indicates the latent space contains disjoint components, violating the unified manifold assumption. A slightly elevated $\beta_0$ under low $W_2$ typically reflects finite-sample artifacts rather than fundamental unification failure.
    
    \item \textbf{Geometric Misalignment:} $\beta_0 \approx 1$ but large $W_2(\hat{\mathbb{P}}_z^{[i]}, \hat{\mathbb{P}}_z^{[j]})$ indicates domains occupy distant regions with insufficient density overlap, violating the open cover requirement (Theorem~\ref{thm:unification}, Condition ii).
    
    \item \textbf{Manifold Collapse:} Low Trust Score $\tau_t$ indicates distinct semantic neighborhoods merge in latent space, signaling violation of the bi-Lipschitz lower bound ($c_1 \to 0$) and thus failure of injectivity (Theorem~\ref{thm:sufficiency}).
    
    \item \textbf{Structural Incoherence:} Low Continuity $\tau_c$ indicates the encoder tears or folds the manifold—neighbors in input space become distant in latent space—suggesting violation of smooth homeomorphism.
    
    \item \textbf{Inconsistent Mapping:} High Alignment Error $\tau_a$ indicates different modalities representing the same entity map to distant latent points, violating cross-modal semantic consistency.
\end{enumerate}

These failure modes form a hierarchical dependency: global unification (items 1--2) is necessary for local preservation (items 3--4) to be meaningful, and both are necessary for semantic coherence (item 5).

\paragraph{Three-Level Verification Framework} 
We verify the existence of a bi-Lipschitz homeomorphism through a hierarchical protocol organized into three verification levels: \emph{global match}, \emph{local match}, and \emph{semantic match}. These levels progressively ensure that the latent space $\mathcal{Z}$ faithfully integrates both the observation space $\mathcal{X}$ and the semantic space $\mathcal{S}$.

\textbf{1. Global Match.} These metrics verify that disparate domains have successfully unified into a single, coherent geometric framework:
\begin{itemize} 
    \item \textbf{Betti Number} ($\beta_0$): Measures the number of connected components in the latent space via persistent homology. Structural integration is achieved when $\beta_0 = 1$, indicating that disparate domains have fused into a single connected manifold rather than remaining as fragmented, disjoint parts. Values of $\beta_0 > 1$ indicate structural fragmentation, while $\beta_0 = 1$ confirms topological unification.
    
    \item \textbf{Sliced Wasserstein Distance} ($W_2$): Measures the geometric distance between probability distributions in latent space using optimal transport theory~\cite{courty2017optimal}. For empirical distributions $\hat{\mathbb{P}}_z^{[i]}$ and $\hat{\mathbb{P}}_z^{[j]}$, small values indicate that latent representations from different domains occupy the same region with substantial density overlap, while large values suggest geometric misalignment. The metric quantifies support coincidence: whether the manifolds induced by different semantic-observation pairs are geometrically coincident.
\end{itemize}

\textbf{2. Local Match.} These metrics verify that the local topology of the observation space is faithfully and smoothly preserved in the latent space:
\begin{itemize} 
    \item \textbf{Trust Score}~\cite{jiang2018trust}: Measures the agreement between the classifier and a modified nearest-neighbor classifier, quantifying whether neighborhoods in the input space are preserved in the embedding space. High trust scores indicate that the encoder maintains local injectivity ($c_1 > 0$), ensuring distinct semantic neighborhoods do not collapse into the same latent region. The metric detects manifold collapse where the bi-Lipschitz lower bound fails.
    
    \item \textbf{Continuity}~\cite{venna2001neighborhood}: Measures whether points that are neighbors in the latent space were also neighbors in the original space, quantifying the smoothness of the learned manifold. High continuity scores indicate that the embedding preserves local structure without introducing spurious proximities or shattering the neighborhood relationships. This metric detects structural incoherence where local variance is high relative to mean distance.
\end{itemize}

\textbf{3. Semantic Match.} This metric verifies the precise synchronization between semantic and observation modalities:
\begin{itemize} 
    \item \textbf{Alignment Error}: Measures the mean distance between embeddings of paired semantic-observation samples from the same underlying entity. For paired data $(x^{(m)}_i, s^{(l)}_i)$ representing the same reality, this metric quantifies whether they map to nearby coordinates in $\mathcal{Z}$:
    \begin{equation}
    \tau_a = \frac{1}{N_{\text{paired}}} \sum_{i=1}^{N_{\text{paired}}} \|E_x^{(m)}(x_i^{(m)}) - E_s^{(l)}(s_i^{(l)})\|_2
    \end{equation}
    Low alignment error indicates consistent cross-modal mapping where the encoder respects the semantic-observation correspondence dictated by the joint data structure. The acceptable threshold depends on whether modalities are homogeneous (same domain) or heterogeneous (cross-domain scenarios).
\end{itemize}

\begin{algorithm}[!htbp]
\caption{Hierarchical Homeomorphism Verification}
\label{alg:homeomorphism}
\begin{algorithmic}[1]
\REQUIRE Integrated latent pool $\mathcal{Z}_{\text{total}} = \bigcup_{k=1}^K \mathcal{Z}^{[k]}$, thresholds $\tau_w, \tau_t, \tau_a, \tau_c$
\REQUIRE Flags: $\textsc{HasPairedModalities}$, $\textsc{RequiresClustering}$

\STATE \textbf{Step 1: Global Match Check}
\IF{$\beta_0(\mathcal{Z}_{\text{total}}) \gg 1$ \AND $W_2(\mathcal{Z}^{[i]}, \mathcal{Z}^{[j]}) > \tau_w$}
    \RETURN \textsc{FAIL}: Structural Fragmentation
\ENDIF
\FOR{$i = 1$ to $K, j = i+1$ to $K$}
    \IF{$W_2(\mathcal{Z}^{[i]}, \mathcal{Z}^{[j]}) > \tau_w$}
        \RETURN \textsc{FAIL}: Geometric Misalignment
    \ENDIF
\ENDFOR

\STATE \textbf{Step 2: Local Match Check}
\FOR{$k = 1$ to $K$}
    \IF{$\text{Trust}_\kappa(\mathcal{Z}^{[k]}) < \tau_t$}
        \RETURN \textsc{FAIL}: Local Manifold Collapse
    \ENDIF
    \IF{$\textsc{RequiresClustering}$ \AND $\text{Cont}_\kappa(\mathcal{Z}^{[k]}) < \tau_c$}
        \RETURN \textsc{FAIL}: Structural Incoherence
    \ENDIF
\ENDFOR

\STATE \textbf{Step 3: Semantic Match Check}
\IF{$\textsc{HasPairedModalities}$}
    \IF{$\text{AlignmentError}(\mathcal{Z}_{\text{total}}) > \tau_a$}
        \RETURN \textsc{FAIL}: Inconsistent Cross-Modal Mapping
    \ENDIF
\ENDIF

\RETURN \textsc{PASS}: $\bigcup_{k=1}^K \mathcal{Z}^{[k]} \cong \mathcal{M}_{\text{univ}}$
\end{algorithmic}
\end{algorithm}

\begin{remark}[Threshold Selection and Numerical Unity]
Thresholds $\tau_w, \tau_t, \tau_a, \tau_c$ are determined empirically through validation on datasets with known compatible and incompatible domain pairs. In practical verification, $\beta_0, W_2,$ and $\tau_c$ must be evaluated in tandem. A manifold may numerically present $\beta_0 > 1$ due to sampling sparsity, yet exhibit $W_2 \approx 0$ and high $\tau_c$, indicating that the domains are geometrically coincident and locally smooth. We define successful unification as the state where domains share a common support that is either path-connected ($\beta_0 = 1$) or geometrically indistinguishable ($W_2$ approaching zero) and locally consistent (high continuity).
\end{remark}

\subsubsection{Theoretical Connection Between ULHM Objectives and Verification Metrics}
The ULHM training objectives (Eq.~\ref{eq:ulhm_loss}) are deliberately designed to enforce the bi-Lipschitz homeomorphism conditions verified by our empirical metrics. We establish the logical chain connecting each training component to its corresponding verification criterion.

\textbf{Preventing Manifold Collapse (Trust Score $\tau_t$).} The reconstruction losses $\mathcal{L}_{\text{recon}}^x$ and $\mathcal{L}_{\text{recon}}^s$ enforce that the encoder must preserve sufficient information to faithfully reconstruct the original inputs. If the encoder were to collapse distinct semantic classes to the same latent region (violating the bi-Lipschitz lower bound $c_1 > 0$), reconstruction quality would degrade as the decoder cannot disambiguate merged representations. This information-preserving pressure maintains local injectivity, which manifests empirically as high Trust scores—semantic neighbors in data space remain semantic neighbors in latent space without spurious mixing.

\textbf{Preserving Local Topology (Continuity $\tau_c$).} The local consistency loss $\mathcal{L}_{\text{local}}$ explicitly minimizes distances between latent codes of $\kappa$-nearest neighbors in data space. This directly implements the continuity requirement of homeomorphism: points close in $\mathcal{M}_x$ must map to points close in $\mathcal{M}_z$. By construction, this objective optimizes precisely for the Continuity metric, ensuring the encoder does not tear the manifold by mapping nearby data points to distant latent regions.

\textbf{Achieving Geometric Unification (Wasserstein Distance $W_2$ and Alignment Error $\tau_a$).} The cross-modal consistency loss $\mathcal{L}_{\text{consist}}$ operates at two complementary levels. The cosine similarity component (weighted by $\lambda_{\text{cos}}$) enforces directional alignment, ensuring different modalities representing the same semantic content point toward the same region in latent space, which directly reduces the distributional distance between domains measured by $W_2$. The Euclidean distance component (weighted by $\lambda_{\text{eucl}}$) further constrains the magnitude alignment, minimizing $\tau_a$ by ensuring paired semantic-observation samples $(s_i, x_i)$ map to nearby coordinates. Together, these components enforce the global connectivity condition (Theorem~\ref{thm:unification}) by ensuring non-trivial overlap $\nu(\mathcal{M}_z^{[i]} \cap \mathcal{M}_z^{[j]}) > 0$ between domain manifolds.

\textbf{Ensuring Topological Unity (Betti Number $\beta_0$).} The combination of reconstruction, consistency, and local preservation losses creates a unified manifold structure. Reconstruction losses prevent the latent space from fragmenting into disconnected task-specific subspaces by requiring all modalities to decode successfully from shared representations. Cross-modal consistency explicitly bridges domains by constraining them to occupy overlapping regions. The perceptual loss $\mathcal{L}_{\text{percep}}$ further reinforces structural coherence by matching high-level feature statistics. These combined pressures ensure the latent space forms a single connected component ($\beta_0 = 1$) rather than disjoint manifolds.

In summary, the ULHM objective function is not an ad hoc combination of losses, but a principled implementation of the bi-Lipschitz homeomorphism criterion: $\mathcal{L}_{\text{recon}}$ enforces the lower bound $c_1 > 0$ (preventing collapse), $\mathcal{L}_{\text{local}}$ enforces the upper bound $c_2 < \infty$ (preventing tearing), and $\mathcal{L}_{\text{consist}}$ enforces global connectivity (enabling unification). Each verification metric subsequently validates whether these theoretical conditions hold empirically on finite samples.

\section{Applications: Sparse Recovery, Transfer, and Zero-Shot Learning}
Having established the ULHM framework for learning latent manifolds and the homeomorphism criterion for verifying when latent manifolds induced by different datasets can be unified, we now demonstrate how verified homeomorphic structure enables three key applications. Each application exploits a distinct property of the universal latent manifold: sparse recovery with semantic priors leverages learned semantic structure to guide reconstruction from incomplete observations, cross-domain integration and transfer unifies heterogeneous domains through verified manifold overlap, and zero-shot learning performs compositional reasoning via continuous manifold interpolation. Critically, all three applications rely on homeomorphism verification from Section~\ref{sec:homeomorphism} to ensure validity without verified homeomorphic structure, these operations would produce unreliable results.

\subsection{Sparse Recovery with Semantic Priors}

We now demonstrate how the ULHM framework enables three key applications through verified homeomorphic structure. We begin with sparse recovery: recovering complete observations from highly incomplete measurements using semantic priors learned during training, even when semantic annotations are unavailable at test time.

\paragraph{Problem Setup}
During training, we observe complete semantic-observation pairs $(\mathbf{s}_i, \mathbf{x}_i) \in \mathcal{D}$ with multiple modalities. For facial image reconstruction on CelebA, the observation modalities consist of: (1) complete ground truth images $x^{(1)} \in \mathbb{R}^{H \times W \times C}$, and (2) sparse masked observations $x^{(2)} = x^{(1)} \odot M$, where $\odot$ denotes element-wise (Hadamard) multiplication and binary mask $M \in \{0,1\}^{H \times W}$ retains only fraction $\rho$ of pixels. The semantic modalities comprise binary facial attributes $s^{(1)} \in \{0,1\}^{K}$ encoding properties such as hair color, presence of glasses, and gender. At deployment, we observe only sparse measurements $\tilde{x}^{(2)}$ without semantic annotations, and must reconstruct the complete image $x^{(1)}$.

\paragraph{Training with the ULHM Framework}
We train the sparse recovery task using the complete ULHM objective (Eq.~\ref{eq:ulhm_loss}). The key mechanism enabling sparse recovery is the cross-modal consistency loss $\mathcal{L}_{\text{consist}}$, which enforces that sparse observations $x^{(2)}$, complete images $x^{(1)}$, and semantic attributes $s^{(1)}$ from the same sample map to nearby regions in $\mathcal{M}_z$. This geometric alignment enables the sparse encoder to implicitly access semantic structure learned from complete observations, even when semantic annotations are unavailable at test time.

\textbf{Semantic Conditioning with Attribute Dropout.} To enable semantic guidance during training while maintaining robustness at test time, we employ attribute dropout: with probability $p_{\text{drop}}$, we zero the attribute embeddings used for conditioning. This forces the encoder to alternate between semantic-guided training (aligning sparse observations with explicit attributes) and observation-only training (extracting semantic structure from visual content alone), ensuring the encoder can leverage implicit semantic geometry at deployment when attributes are unavailable.

\paragraph{Test-Time Recovery via Amortized Inference}
At deployment, given only sparse observations $\tilde{x}^{(2)}$ and mask $M$ without semantic annotations, we reconstruct the complete image through direct amortized inference:
\begin{equation}
x^* = D_\theta^{(1)}(E_x^{(2)}([\tilde{x}^{(2)}, M]))
\end{equation}
where the sparse encoder $E_x^{(2)}$ maps incomplete observations to the latent manifold $\mathcal{M}_z$, and the decoder $D_\theta^{(1)}$ reconstructs the complete image modality. If the latent manifolds are homeomorphic (verified via Algorithm~\ref{alg:homeomorphism}), semantic structure learned during training is embedded as geometric constraints in $\mathcal{M}_z$, enabling recovery from incomplete observations without explicit semantic input. 

\paragraph{Dependence on Verified Homeomorphism}
The effectiveness of this approach depends critically on verified homeomorphism. If $\text{Trust}_\kappa$ or $\text{Cont}_\kappa$ falls below threshold, the latent manifold does not faithfully preserve data topology, and sparse observations may map to regions where semantic structure is unreliable, causing semantically incoherent reconstructions. Conversely, when homeomorphism verification passes, the bi-Lipschitz property (Theorem~\ref{thm:sufficiency}) guarantees that distinct semantic states occupy geometrically separated regions in $\mathcal{M}_z$, enabling reliable reconstruction from sparse measurements.   

\subsection{Cross-Domain Integration and Transfer}

A key advantage of the universal latent manifold is enabling classifier transfer across heterogeneous datasets without retraining. Consider two datasets: $\mathcal{D}^{[1]} = \{(\mathbf{s}_i^{[1]}, \mathbf{x}_i^{[1]})\}_{i=1}^{N_1}$ and $\mathcal{D}^{[2]} = \{(\mathbf{s}_j^{[2]}, \mathbf{x}_j^{[2]})\}_{j=1}^{N_2}$ from different domains but sharing similar semantic structure (e.g., MNIST handwritten digits and Fashion-MNIST clothing items, both with shared class labels). The central question is: can a classifier trained solely on the first dataset directly classify samples from the second dataset without any adaptation or retraining?

The homeomorphism criterion from Section~\ref{sec:homeomorphism} provides the answer. We first apply Algorithm~\ref{alg:homeomorphism} to verify three conditions: (1) individual homeomorphism through $\text{Trust}_\kappa$ and $\text{Cont}_\kappa$ for both datasets to confirm each preserves its topology in latent space, (2) manifold overlap through $W_2(\hat{P}_z^{[1]}, \hat{P}_z^{[2]})$ to verify the latent manifolds induced by different datasets occupy the same region of $\mathcal{M}_z$, and (3) alignment consistency to verify consistent semantic-observation mapping. When all three conditions pass, cross-domain transfer becomes theoretically grounded and empirically reliable.

\paragraph{Three-Stage Training Protocol}
Unlike methods that jointly train all components, our framework employs a three-stage protocol that ensures fair transfer evaluation by strictly separating domain-specific learning from universal alignment. For cross-domain transfer, we treat each dataset as a unified entity with domain-specific encoder and decoder networks.

\textbf{Stage 1: Domain-Specific Encoder Pre-training.} We independently pre-train domain-specific encoders $E^{[k]}: \mathcal{X}^{[k]} \to \mathcal{Z}^{[k]}$ and decoders $D^{[k]}: \mathcal{Z}^{[k]} \to \mathcal{X}^{[k]}$ for each dataset using the complete ULHM objective (Eq.~\ref{eq:ulhm_loss}) applied within each domain independently:
\begin{equation}
\mathcal{L}_{\text{pretrain}}^{[k]} = \mathcal{L}_{\text{ULHM}}\big|_{\mathcal{D}^{[k]}}
\end{equation}
where $\tilde{x}_i^{[k]} = x_i^{[k]} \odot M_i$ represents sparse observations with mask $M_i$ retaining fraction $\rho$ of measurements, and $x_i^{[k]}$ is the complete ground truth. This stage learns semantically-enriched domain-specific representations without cross-domain interaction, establishing independent baseline representations for each dataset.

\textbf{Stage 2: Domain-Invariant Projection with Alignment.} After freezing the pre-trained encoders, we train projection networks $\Pi^{[k]}: \mathcal{Z}^{[k]} \to \mathcal{M}_z$ that map domain-specific latent codes to the universal manifold:
\begin{equation}
z_{\text{univ},i}^{[k]} = \Pi^{[k]}(E^{[k]}([\tilde{x}_i^{[k]}, M_i]))
\end{equation}
The projection networks are trained jointly using a composite objective that combines reconstruction, contrastive alignment~\cite{chen2020simple}, and centroid alignment~\cite{wen2016discriminative}:
\begin{align}
\mathcal{L}_{\text{align}} = &\sum_{k=1}^K \sum_{i=1}^{N_k} \|D^{[k]}(z_{\text{univ},i}^{[k]}) - x_i^{[k]}\|_2^2  + \lambda_{\text{cont}} \mathcal{L}_{\text{contrastive}} + \lambda_{\text{cent}} \mathcal{L}_{\text{centroid}}
\end{align}
where the contrastive loss encourages samples with matching labels from different domains to cluster together:
\begin{equation}
\mathcal{L}_{\text{contrastive}} = -\sum_{i,j} \mathbbm{1}[y_i^{[k]} = y_j^{[k']}] \log \frac{\exp(\text{sim}(z_{\text{univ},i}^{[k]}, z_{\text{univ},j}^{[k']})/\tau)}{\sum_{n} \exp(\text{sim}(z_{\text{univ},i}^{[k]}, z_{\text{univ},n})/\tau)}
\end{equation}
where $\mathbbm{1}[\cdot]$ is the indicator function, $\text{sim}(\cdot,\cdot)$ denotes cosine similarity, and $\tau  $ is the temperature parameter. The centroid loss enforces that class centroids $\mu_c^{[k]} = \frac{1}{|\mathcal{I}_c^{[k]}|}\sum_{i \in \mathcal{I}_c^{[k]}} z_{\text{univ},i}^{[k]}$, where $\mathcal{I}_c^{[k]} = \{i: y_i^{[k]} = c\}$ is the set of sample indices with class label $c$ in dataset $k$, coincide across domains:
\begin{equation}
\mathcal{L}_{\text{centroid}} = \sum_{c=0}^{C-1} \sum_{k,k'=1, k<k'}^{K} \|\mu_c^{[k]} - \mu_c^{[k']}\|_2^2
\end{equation}
with $\mathcal{I}_c^{[k]} = \{i: y_i^{[k]} = c\}$ denoting sample indices with class label $c$ in domain $k$, $\text{sim}(\cdot,\cdot)$ as cosine similarity, and $\tau$ as temperature. Critically, no classifier is trained during this stage—the alignment losses operate solely on the learned universal manifold structure, ensuring that semantic correspondence emerges from geometric constraints rather than supervised classification pressure.

\textbf{Stage 3: Single-Domain Classifier Training.} After the universal manifold is established, we train a classifier $C_\theta: \mathcal{M}_z \to \mathcal{Y}$ exclusively on one domain (e.g., MNIST) while keeping all encoders and projections frozen:
\begin{equation}
\mathcal{L}_{\text{class}}^{[1]} = \sum_{i=1}^{N_1} \mathcal{L}_{\text{CE}}(C_\theta(z_{\text{univ},i}^{[1]}), y_i^{[1]})
\end{equation}
where $z_{\text{univ},i}^{[1]} = \Pi^{[1]}(E^{[1]}([\tilde{x}_i^{[1]}, M_i]))$ and $\mathcal{L}_{\text{CE}}$ is the cross-entropy loss. This single-domain training protocol is essential for fair transfer evaluation: the classifier never observes samples from the target domain, ensuring that any successful transfer truly validates the learned homeomorphic structure rather than memorized cross-domain patterns.

\paragraph{Zero-Shot Cross-Domain Transfer}
At test time, the trained classifier directly operates on the target domain (e.g., Fashion-MNIST) without any adaptation. For a test sample $\tilde{x}^{[2]}$ from the target domain:
\begin{equation}
\hat{y} = \arg\max_c C_\theta(z_{\text{univ}}^{[2]})_c, \quad z_{\text{univ}}^{[2]} = \Pi^{[2]}(E^{[2]}([\tilde{x}^{[2]}, M]))
\end{equation}
The homeomorphic structure guarantees successful transfer because: (1) the projection $\Pi^{[2]}$ maps target domain samples to the same universal manifold $\mathcal{M}_z$ where the classifier was trained, and (2) samples with the same semantic label from different domains occupy nearby regions due to verified alignment, enabling the decision boundaries learned on the source domain to generalize to the target domain.

This zero-shot capability eliminates the need for: (1) labeled data from the target domain, (2) fine-tuning or domain adaptation procedures, and (3) separate classifiers per domain. A single classifier trained on one domain transfers across all verified homeomorphic datasets.

\paragraph{Dependence on Verified Homeomorphism}
The zero-shot transfer capability depends critically on verified homeomorphism. When homeomorphism verification passes (Algorithm~\ref{alg:homeomorphism}), the classifier successfully transfers across domains with minimal accuracy degradation. If verification fails, cross-domain transfer produces unreliable predictions. Our experiments in Section~\ref{sec:experimental} demonstrate that with verified homeomorphic structure, a classifier trained exclusively on MNIST achieves substantial accuracy on Fashion-MNIST without ever observing Fashion-MNIST samples during training, and vice versa, validating that the homeomorphism criterion correctly predicts when zero-shot transfer will succeed.

\subsection{Zero-Shot Learning via Manifold Learning}

Having demonstrated sparse recovery (Section 3.1) and cross-domain transfer (Section 3.2), we now show how the ULHM framework enables zero-shot classification through topological preservation. Unlike the previous applications that rely on cross-modal alignment or domain-invariant projections, zero-shot learning exploits the local consistency loss $\mathcal{L}_{\text{local}}$ to ensure unseen classes naturally cluster in geometrically separated regions of $\mathcal{M}_z$.

\paragraph{Problem Setup}
During training, we observe observation-label pairs $(\mathbf{x}_i, \mathbf{s}_i) \in \mathcal{D}$ where the semantic modality consists of class labels $s^{(l)} \in \{0, 1, \ldots, C-1\}$. We deliberately withhold certain classes from classifier training by partitioning the label space into seen classes $\mathcal{Y}^s = \{0, \ldots, C_s-1\}$ and unseen classes $\mathcal{Y}^u = \{C_s, \ldots, C-1\}$. Critically, we train encoders $E_x^{(m)}$ and decoders $D_\theta^{(m)}$ on \emph{all} classes using the complete ULHM objective (Eq.~\ref{eq:ulhm_loss}), while the auxiliary classifier $C_\theta: \mathcal{M}_z \to \mathbb{R}^{C_s}$ sees only the seen classes subset. At test time, we face a zero-shot scenario: can the framework correctly predict classes the classifier has never been trained on?

\paragraph{Training with the ULHM Framework}
We apply the ULHM objective (Eq.~\ref{eq:ulhm_loss}) with class labels as the only semantic modality. For unseen classes, we use only the observation reconstruction term $\mathcal{L}_{\text{recon}}^x$ and topological preservation loss $\mathcal{L}_{\text{local}}$, omitting semantic reconstruction $\mathcal{L}_{\text{recon}}^s$ and cross-modal consistency $\mathcal{L}_{\text{consist}}$ since we deliberately withhold their labels:
\begin{equation}
\mathcal{L}_{\text{zero-shot}} = \underbrace{\sum_{i: s_i^{(l)} \in \mathcal{Y}^s} \mathcal{L}_{\text{ULHM}}\big|_i}_{\text{seen classes: full ULHM}} + \underbrace{\sum_{i: s_i^{(l)} \in \mathcal{Y}^u} (\mathcal{L}_{\text{recon}}^x + \lambda_\ell \mathcal{L}_{\text{local}} + \lambda_p \mathcal{L}_{\text{percep}})\big|_i}_{\text{unseen classes: observation-only}}
\end{equation}
Additionally, we train an auxiliary classifier only on seen classes:
\begin{equation}
\mathcal{L}_{\text{classifier}} = \sum_{i: s_i^{(l)} \in \mathcal{Y}^s} \mathcal{L}_{\text{CE}}(C_\theta(E_x^{(m)}(x_i^{(m)})), s_i^{(l)})
\end{equation}
The key mechanism enabling zero-shot transfer is the topological preservation loss $\mathcal{L}_{\text{local}}$, which enforces that visually similar samples cluster together regardless of whether their labels are used for classifier training. For unseen classes, this clustering emerges purely from visual similarity preserved through reconstruction and neighborhood structure, ensuring they occupy well-defined, geometrically separated regions in $\mathcal{M}_z$ even without classifier supervision.

\paragraph{Nearest-Centroid Classification in Latent Space}
For a test sample $x_{\text{test}}^{(m)}$ from an unseen class, we perform nearest-centroid classification directly in the learned latent manifold. For each class $c \in \{0, \ldots, C-1\}$ (including unseen classes), we compute the centroid using all available training samples:
\begin{equation}
\mu_c = \frac{1}{|\mathcal{I}_c|} \sum_{i \in \mathcal{I}_c} E_x^{(m)}(x_i^{(m)}; \phi)
\end{equation}
where $\mathcal{I}_c = \{i : s_i^{(l)} = c\}$ is the set of training sample indices with class label $c$. Crucially, we can compute centroids for unseen classes because the encoders processed these samples during training through the ULHM reconstruction objectives, even though the classifier never saw their labels.

Classification proceeds by finding the nearest centroid in latent space:
\begin{equation}
\hat{c} = \arg\min_{c \in \{0,\ldots,C-1\}} \|E_x^{(m)}(x_{\text{test}}^{(m)}; \phi) - \mu_c\|_2
\end{equation}
For generalized zero-shot learning (GZSL), the search space includes both seen and unseen classes. For conventional zero-shot learning (ZSL), the search space is restricted to $\mathcal{Y}^u$ only.

\paragraph{Theoretical Foundation: Homeomorphism Enables Zero-Shot Transfer} 
The zero-shot capability fundamentally relies on the homeomorphic structure established in Section~\ref{sec:homeomorphism}. By Theorem~\ref{thm:sufficiency}, the encoder $E_x^{(m)}$ satisfies the bi-Lipschitz condition with $c_1 > 0$, ensuring it is a homeomorphism onto its image. This topological preservation guarantees three critical properties:

\begin{enumerate}
\item \textbf{Injectivity ($c_1 > 0$):} Distinct visual samples map to distinct latent codes, preventing manifold collapse. This ensures that unseen classes occupy well-defined, separable regions in $\mathcal{M}_z$ even without classifier supervision.

\item \textbf{Local continuity (upper bound $c_2$):} The local consistency loss $\mathcal{L}_{\text{local}}$ enforces that samples from the same class cluster together in latent space. For unseen classes, this clustering emerges purely from geometric similarity of visual observations, as the encoder preserves neighborhood structure through topology preservation.

\item \textbf{Global structure preservation:} The reconstruction objectives $\mathcal{L}_{\text{recon}}^x$ and $\mathcal{L}_{\text{recon}}^s$ ensure the latent manifold captures discriminative visual features. Combined with homeomorphism, this guarantees that class centroids $\{\mu_c\}_{c \in \mathcal{Y}^u}$ for unseen classes occupy distinct regions geometrically separated from seen class centroids $\{\mu_c\}_{c \in \mathcal{Y}^s}$.
\end{enumerate}

\paragraph{Dependence on Topological Structure}  Nearest-centroid classification succeeds in the zero-shot regime because the homeomorphic encoder maps the \emph{entire} data manifold $\mathcal{M}_{\mathcal{X}}$ (including unseen classes) to the latent manifold $\mathcal{M}_z$ while preserving topological invariants. The classifier $C_\theta$ trained only on seen classes learns decision boundaries in a subspace of $\mathcal{M}_z$, but the homeomorphic structure ensures that geometric relationships extend globally. Unseen class samples naturally cluster at their correct centroids through preserved topology, enabling classification without supervision.
If topology preservation fails (e.g., manifold collapse where $c_1 \to 0$, or tearing where continuity is violated), the Trust metric (Proposition~\ref{prop:bridge}) drops below threshold, indicating that class boundaries become unreliable. In such cases, zero-shot classification produces arbitrary predictions, as unseen class samples no longer cluster coherently in latent space.

\section{Experimental Validation}
\label{sec:experimental}
We validate the ULHM framework across three experimental domains corresponding to the applications in Section III: (1) sparse recovery with semantic priors using MNIST digits and CelebA facial images with binary attributes, (2) cross-domain integration and transfer using MNIST-Fashion-MNIST pairs, and (3) zero-shot classification on MNIST, Fashion-MNIST, and CIFAR-10. For each application, homeomorphism preservation is enforced as an intrinsic training objective through the   consistency loss   and verified post-training via Algorithm~\ref{alg:homeomorphism} to ensure the learned representations satisfy the theoretical requirements for valid unification and transfer.

\subsection{Implementation Details}

\textbf{Latent Dimensionality.} We set latent dimension $d=512$ for MNIST/Fashion-MNIST to support universal classifier learning and cross-domain transfer, $d=1024$ for CelebA to accommodate high-dimensional attribute embeddings and reconstruction quality, and $d=512$ for CIFAR-10 to balance capacity with computational efficiency.

\textbf{Architecture.} We implement encoders $E_x^{(m)}$ and decoders $D_\theta^{(m)}$ using convolutional neural networks for image modalities. For MNIST and Fashion-MNIST (28×28 grayscale), encoders consist of 4 convolutional layers (32, 64, 128, 256 channels) with 3×3 kernels, stride 2, and LeakyReLU activations, followed by fully connected layers projecting to $z \in \mathbb{R}^d$. Decoders mirror this architecture with transposed convolutions. For CelebA (64×64 RGB), we use ResNet-based encoders with Adaptive Instance Normalization (AdaIN) blocks for semantic conditioning, incorporating attribute embedding networks that project binary attributes into continuous embedding space before modulating feature statistics. For CIFAR-10 (32×32 RGB), we use Wide Residual Networks \cite{zagoruyko2016wide} as the backbone for zero-shot learning, adapted with 4 convolutional layers to the spatial resolution. The auxiliary classifier $C_\theta$ for class labels uses either a 3-layer MLP with hidden dimensions [512, 256, num\_classes] for MNIST/Fashion-MNIST, or a spatial classifier with convolutional feature extraction followed by adaptive pooling for cross-domain transfer tasks.

\textbf{Datasets.} We evaluate our framework on four benchmark datasets. MNIST contains 60,000 training and 10,000 test grayscale images ($28\times28$) of handwritten digits across 10 classes. Fashion-MNIST follows the same structure with clothing items. CelebA contains 162,770 training and 19,962 test facial images ($64\times64$ after center cropping and resizing) with 40 binary attributes including hair color, accessories, and facial features. CIFAR-10 contains 50,000 training and 10,000 test RGB images ($32\times32$) across 10 natural object categories. All images are normalized to $[0,1]$ range.

\textbf{Evaluation Metrics.} We report Mean Squared Error (MSE) for reconstruction quality and accuracy for classification tasks. 
For homeomorphism verification, we compute Trust Score~\cite{jiang2018trust}, Continuity~\cite{venna2001neighborhood}, Sliced Wasserstein-2 distance, Betti number $\beta_0$, and Alignment Error as defined in Section~\ref{sec:homeomorphism}, with neighborhood size $\kappa = 5$. 

We set verification thresholds as follows: Trust $\tau_t \geq 0.80$ following the original trust score formulation~\cite{jiang2018trust}, which demonstrates that scores above 0.80 reliably indicate preserved neighborhood structure in manifold learning. Continuity $\tau_c \geq 0.70$ aligns with empirical observations in manifold learning literature~\cite{venna2001neighborhood} where values above 0.70 indicate acceptable local structure preservation. For Betti number, structural integration requires $\beta_0 = 1$, indicating a single connected component. Sliced Wasserstein distance $\tau_w \leq 0.30$ is set based on domain adaptation studies~\cite{courty2017optimal}, where inter-domain distances below 0.3 indicate successful alignment relative to typical intra-domain variation. Alignment Error $\tau_a \leq 0.30$ accommodates the natural variation in cross-domain heterogeneous scenarios while ensuring meaningful semantic-observation correspondence, as opposed to stricter thresholds ($\tau_a \leq 0.15$) suitable for single-domain paired modalities.

\textbf{Baselines.} We evaluate our framework against a comprehensive suite of classical benchmarks and state-of-the-art methods. 
\begin{itemize}
 
    \item For \textit{sparse recovery}, we compare against the following leading methods: Attention U-Net \cite{oktay2018attention}, Dense U-Net \cite{li2018h}, DANet \cite{fu2019dual}, Residual Attention \cite{wang2017residual}, CBAM U-Net \cite{woo2018cbam}, ResUNet-PSP \cite{diakogiannis2020resunet}, and ResUNet \cite{zhang2018road}.
 
\item For \textit{cross-domain transfer learning}, we compare against several representative and state-of-the-art domain adaptation and transfer learning methods, including Partial Convolution (PConv) \cite{liu2018partialconv}, Maximum Classifier Discrepancy (MCD) \cite{saito2018mcd}, Conditional Domain Self-supervised Learning (CDSL) \cite{kim2020crossdomain}, Minimum Class Confusion (MCC) \cite{jin2020mcc}, Conditional Domain Adversarial Networks (CDAN) \cite{long2018cdan}, Deep Correlation Alignment (CORAL) \cite{sun2016deep}, and Prototypical Networks (ProtoNet) \cite{snell2017protonet}.

 \item For \textit{zero-shot learning}, we compare against widely used state-of-the-art baselines including Contrastive Learning (SimCLR) \cite{chen2020simple}, Prototypical Networks \cite{snell2017protonet}, Triplet Networks \cite{hoffer2015deep}, Variational Autoencoders (VAE) \cite{kingma2014auto}, Maximum Classifier Discrepancy (MCD) \cite{saito2018mcd}, Deep Correlation Alignment (CORAL) \cite{sun2016deep}, Mixup Augmentation \cite{zhang2018mixup}, and Self-supervised Rotation Prediction \cite{gidaris2018unsupervised}.

\end{itemize}

\subsection{Image Sparse Recovery}
We evaluate the ULHM framework's ability to recover complete observations from sparse measurements using semantic priors learned during training. During training, models observe complete semantic-observation pairs $(s_i, x_i)$ at a single sparsity level ($\rho = 0.10$). At test time, we evaluate recovery performance across multiple sparsity levels $\rho \in \{0.10, 0.15, 0.20, 0.25\}$ by providing only incomplete observations $\tilde{x}^{(m)} = x^{(m)} \odot M$, where mask $M$ retains only fraction $\rho$ of measurements, with no semantic guidance.
\begin{figure}[!htb]
    \centering
    \includegraphics[width=0.9\linewidth]{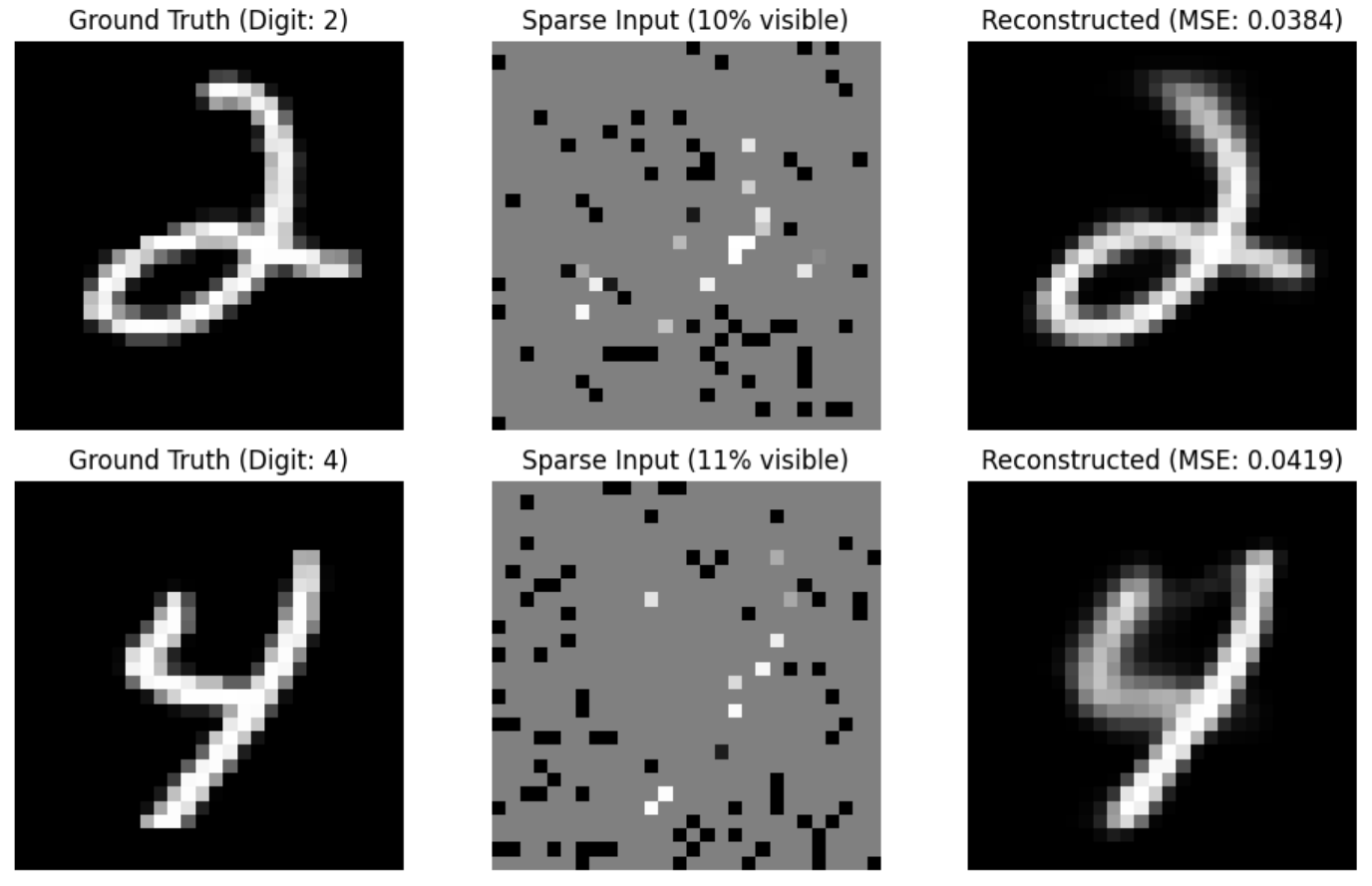}
    \caption{Sparse recovery on the MNIST digits. It shows ground truth images, heavily masked 10\%--11\% inputs, and the successful corresponding reconstructed image with low MSE. }
    \label{fig:sample_mnist}
    \vspace{-0.3cm}
\end{figure}

\subsubsection{MNIST}
\textbf{Setup.} MNIST digits (28×28 = 784 pixels) are paired with semantic information consisting of: (1) complete ground truth images $x^{(1)} \in \mathbb{R}^{28 \times 28}$ and (2) class labels $s^{(1)} \in \{0,1,...,9\}$. The framework is trained exclusively on $\rho = 0.10$ sparsity, where sparse observations $x^{(2)}$ retain only 10\% of pixels through random masking with uniform spatial distribution as shown in Figure ~\ref{fig:sample_mnist}. At test time, we evaluate generalization by testing on sparsity levels ranging from 0.10 to 0.25 without any semantic information provided.

\begin{table}[!htpb]
\centering
\caption{Homeomorphism verification metrics of MNIST ($\kappa=5$, Cosine distance)}
\label{tab:mnist_homeo}
\begin{tabular}{llcccc}
\hline
\textbf{Level} & \textbf{Metric} & ${\rho=0.10}$ & ${\rho=0.15}$ & ${\rho=0.20}$ & ${\rho=0.25}$ \\
\hline
\multirow{2}{*}{Global Match} & Betti number $\beta_0$ & 1 & 1 & 1 & 1 \\
 & Sliced $W_2$ $\tau_w$  & 0.014 & 0.019 & 0.024 & 0.027 \\
\hline
\multirow{2}{*}{Local Match} & Trust $\tau_t$ & 0.800 & 0.902 & 0.937 & 0.953 \\
 & Continuity $\tau_c$ & 0.906 & 0.899 & 0.890 & 0.882 \\
\hline
Semantic Match & Alignment $\tau_a$ & 0.027 & 0.030 & 0.034 & 0.038 \\
\hline
\end{tabular}
\end{table}

\textbf{Homeomorphism Verification.} We verify that the learned latent manifold preserves topological structure across different sparsity regimes using four metrics computed over the complete dataset (60,000 training and 10,000 test samples). Table~\ref{tab:mnist_homeo} presents results with neighborhood size $\kappa=5$. The 0-th Betti number $\beta_0 = 1$ across all sparsity levels confirms that the latent space forms one connected manifold, validating that sparse and full encoders learn a unified geometric structure rather than separate components. Trust scores increase monotonically from 0.80 at $\rho=0.10$ to 0.953 at $\rho=0.25$, demonstrating that higher measurement density strengthens semantic clustering. The negligible train-test difference (average $\Delta = 0.004$) and small Wasserstein distance ($W_2 \approx 0.03$, one order of magnitude below threshold 0.3) confirm robust generalization with no domain shift. Alignment error increases from 0.027 to 0.038, reflecting the framework's adaptive encoding: at low sparsity, encoders converge to canonical class representations (leveraging semantic priors), while at higher sparsity, encoders capture instance-specific details. This behavior, combined with $W_2 \approx 0.03$, demonstrates encoding at different granularities within the same manifold rather than mapping to separate manifolds. All sparsity levels satisfy the connectivity ($\beta_0=1$), trust ($\geq 0.80$), and co-location ($W_2 \ll 0.3$) criteria, conclusively demonstrating one latent manifold with preserved structure.

\begin{table}[!htbp]
\centering
\caption{MNIST Sparse Recovery Performance Across Sparsity Levels}
\label{tab:mnist_sparse}
\begin{tabular}{lcccc}
\hline
Method & $\rho = 0.10$ & $\rho = 0.15$ & $\rho = 0.20$ & $\rho = 0.25$ \\
\hline
UNet \cite{ronneberger2015u} & 0.084 & 0.051 & 0.035 & 0.027 \\
ResNet \cite{he2016deep} & 0.074 & 0.044 & 0.031 & 0.023 \\
DenseNet \cite{huang2017densely} & 0.092 & 0.055 & 0.037 & 0.027 \\
AttentionUNet \cite{oktay2018attention} & 0.084 & 0.051 & 0.035 & 0.026 \\
VAE \cite{kingma2014auto} & 0.097 & 0.077 & 0.070 & 0.068 \\
\hline
\textbf{Ours} & \textbf{0.054} & \textbf{0.035} & \textbf{0.027} & \textbf{0.021} \\
\hline
\end{tabular}
% \vspace{2pt}
% \footnotesize Baselines report PSNR ($\uparrow$); {\color{red} Ours reports MSE ($\downarrow$) with std: 0.031, 0.020, 0.014, 0.011.}
\end{table}

\textbf{Recovery Performance.} Table~\ref{tab:mnist_sparse} presents reconstruction quality measured by mean squared error (MSE). Despite training exclusively on $\rho=0.10$ data, ULHM achieves superior performance across all sparsity levels. At the training sparsity ($\rho=0.10$), ULHM achieves MSE of 0.054, outperforming the best baseline ResNet (0.074) by 27.0\%. This advantage becomes more pronounced at higher sparsities: at $\rho=0.15$, ULHM (0.035) outperforms ResNet (0.044) by 20.5\%; at $\rho=0.20$, ULHM (0.027) outperforms ResNet (0.031) by 12.9\%; and at $\rho=0.25$, ULHM (0.021) outperforms ResNet (0.023) by 8.7\%. Notably, ULHM's performance at $\rho=0.25$ (0.021) is 61.1\% better than its own performance at the training sparsity $\rho=0.10$ (0.054), demonstrating effective generalization beyond the training distribution. The monotonic decrease in MSE as more observations become available validates that semantic priors learned during training provide increasingly stronger geometric constraints for reconstruction.
\begin{figure}[!htbp]
    \centering
    \includegraphics[width=0.80\linewidth]{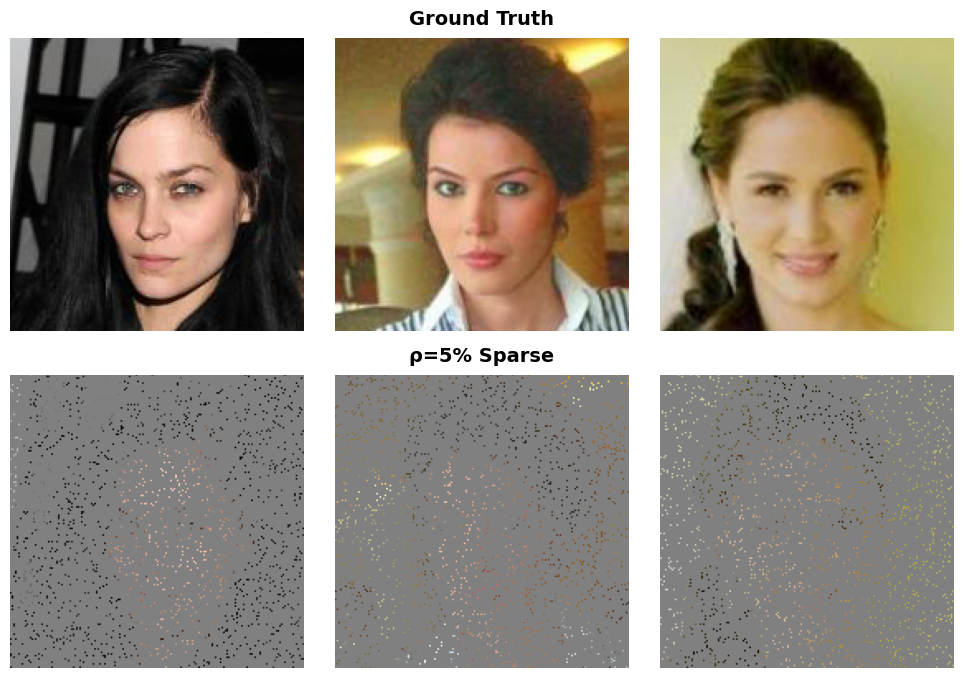}
    \caption{Illustrating the sparse CelebA-style image recovery settings with ground truth and random 5\% sparsity.}
    \label{fig:celeba_settings}
    \vspace{-0.4cm}
\end{figure}
%%%

 \begin{figure*}[!htb]
    \centering
    \includegraphics[width=0.95\linewidth]{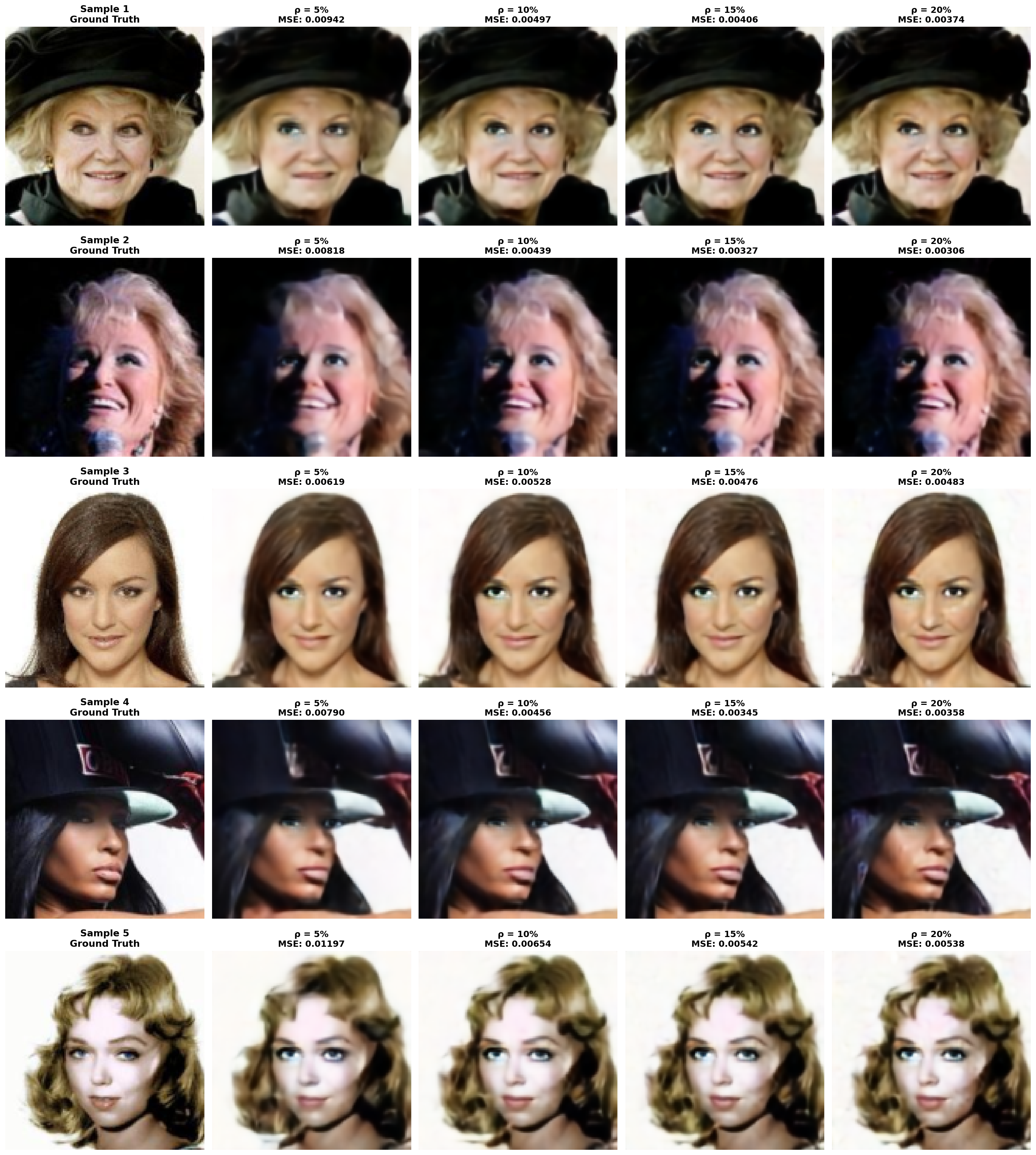}
    \caption{Sparse face reconstructions under increasing observation rates $\rho (5\%-20\%)$, showing consistent visual improvement, and decreasing MSE. }
    \label{fig:celeba}
    \vspace{-0.6cm}
\end{figure*}

\subsubsection{CelebA}
\textbf{Setup.} CelebA facial images (64×64×3 = 12,288 values) are paired with 40 binary facial attributes including \textit{Black\_Hair}, \textit{Blond\_Hair}, \textit{Eyeglasses}, \textit{Male}, \textit{Smiling}, \textit{Young}, etc. We test spatially structured masking patterns: (1) random pixel masking  (Fig. \ref{fig:celeba_settings}), (2) block masking (removing 16×16 patches), and (3) inpainting (masking central 32×32 region).

\begin{table}[!htpb]
\centering
\caption{Topological Unification Verification - CelebA ($\kappa=5$, Cosine distance)}
\label{tab:celeba_unification}
\begin{tabular}{llccccc}
\hline
\textbf{Level} & \textbf{Metric} & $\rho=0.05$ & $\rho=0.08$ & $\rho=0.10$ & $\rho=0.12$ & $\rho=0.15$ \\
\hline
\multirow{2}{*}{Global Match} & Betti number $\beta_0$ & 1 & 1 & 1 & 1 & 1 \\
 & Sliced $W_2$ $\tau_w$  & 0.006 & 0.008 & 0.011 & 0.012 & 0.013 \\
\hline
\multirow{2}{*}{Local Match} & Trust $\tau_t$  & 0.877 & 0.873 & 0.871 & 0.868 & 0.866 \\
 & Continuity $\tau_c$  & 0.888 & 0.889 & 0.885 & 0.886 & 0.879 \\
\hline
Semantic Match & Alignment $\tau_a$  & 0.043 & 0.041 & 0.043 & 0.046 & 0.055 \\
\hline
\end{tabular}
\end{table}

\textbf{Homeomorphism Verification.} Table~\ref{tab:celeba_unification} confirms homeomorphic structure is preserved across all sparsity levels with Trust$_{5} \geq 0.866 > 0.80$ and Cont$_{5} \geq 0.879 > 0.70$, meeting verification thresholds. The Betti number $\beta_0 = 1$ across all sparsity levels confirms that the latent space forms a single connected manifold, validating structural integration. Sliced $W_2$ distances remain extremely small ($W_2 \leq 0.013 \ll 0.30$), demonstrating that sparse and complete encoders map to geometrically coincident regions with substantial density overlap. Alignment error increases modestly from 0.043 to 0.055 as sparsity increases, reflecting adaptive encoding at different granularities while maintaining manifold coherence. The consistent satisfaction of all verification criteria (connectivity with $\beta_0=1$, local preservation with Trust $> 0.80$ and Continuity $> 0.80$, and geometric co-location with $W_2 \ll 0.30$) conclusively demonstrates that CelebA sparse recovery operates on a unified homeomorphic latent manifold.

\begin{table}[!htpb]
\label{tab:celeba_sparse}
\centering
\caption{CelebA sparse recovery with Gaussian noise ($\sigma=0.6$)}
\label{tab:celeba_noisy_performance}
\begin{tabular}{lccccc}
\hline
\textbf{Method} & ${\rho=0.05}$ & ${\rho=0.08}$ & ${\rho=0.10}$ & ${\rho=0.12}$ & ${\rho=0.15}$ \\
\hline
Attention U-Net \cite{oktay2018attention} & 0.0406 & 0.0404 & 0.0401 & 0.0396 & 0.0388 \\
Dense U-Net \cite{li2018h} & 0.0342 & 0.0334 & 0.0328 & 0.0321 & 0.0312 \\
DANet \cite{fu2019dual} & 0.0387 & 0.0364 & 0.0348 & 0.0332 & 0.0309 \\
Res-Attention \cite{wang2017residual} & \textbf{0.0325} & 0.0313 & 0.0307 & 0.0301 & 0.0292 \\
CBAM U-Net \cite{woo2018cbam} & 0.0348 & 0.0322 & 0.0307 & 0.0293 & 0.0273 \\
ResUNet-PSP \cite{diakogiannis2020resunet} & 0.0336 & 0.0319 & 0.0313 & 0.0306 & 0.0297 \\
ResUNet \cite{zhang2018road} & 0.0360 & 0.0342 & 0.0330 & 0.0317 & 0.0297 \\
\hline
\textbf{ ULHM (Ours)} &  {0.0353} & \textbf{0.0305} & \textbf{0.0280} & \textbf{0.0262} & \textbf{0.0247} \\
\hline
\end{tabular}
\end{table}

 \begin{figure*}[!htb]
    \centering
    \includegraphics[width=1.0\linewidth]{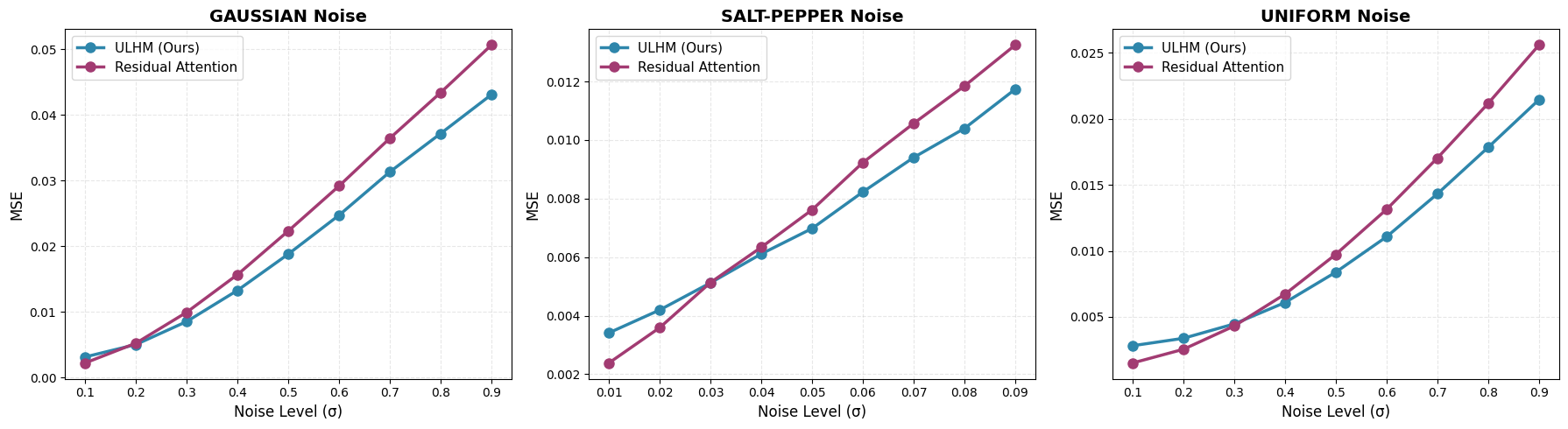}
    \caption{Mean Squared Error (MSE) under increasing Gaussian, Salt-Pepper, and Uniform noise. ULHM achieved lower than the Residual Attention baseline, demonstrating superior robustness.}
    \label{fig:mse_noise}
    \vspace{-0.6cm}
\end{figure*}

\textbf{Sparse Recovery Results.} Table~\ref{tab:celeba_noisy_performance} shows reconstruction performance under Gaussian noise ($\sigma=0.6$) across different sparsity levels. ULHM demonstrates consistent improvement as sparsity increases, achieving MSE of 0.0280 at $\rho=0.10$, outperforming all baselines including Res-Attention (0.0307) and CBAM U-Net (0.0307). At higher sparsity levels, the gap widens significantly: at $\rho=0.15$, ULHM achieves MSE of 0.0247 compared to the best baseline CBAM U-Net at 0.0273, demonstrating a 9.5\% improvement. Notably, while Res-Attention performs best among baselines at $\rho=0.05$ (0.0325 vs. ULHM's 0.0353), ULHM's performance scales more effectively with increased measurements, validating that semantic structure learned during training provides stronger geometric constraints as more observations become available.

\textbf{Robustness against noise.} Figure \ref{fig:mse_noise} depicts robustness of our method against different noise models. This was to evaluate how the reconstruction or prediction error increases with the input noise severity. ULHM consistently outperforms Residual Attention across Gaussian, salt-and-pepper and uniform noise, reducing MSE by 15-20\% at high noise levels (e.g., 0.043 vs. 0.051 for Gaussian $\sigma =0.9$, 0.012 vs. 0.014 for salt-and-pepper  $\sigma=0.09$, and 0.021 vs. 0.026 for uniform $\sigma=0.9$). It resembles the ULHM's capability of handling any noise and impulsive pixel corruption.

\subsection{Cross-Domain Integration and Transfer}
We validate the framework's ability to integrate heterogeneous datasets through verified homeomorphic structure. We test MNIST-Fashion-MNIST integration where shared class labels $\{0,1,...,9\}$ apply to fundamentally different visual domains (handwritten digits vs. clothing items), demonstrating zero-shot classifier transfer across domains without retraining.

\subsubsection{Verification}
\textbf{MNIST and Fashion-MNIST.} We train ULHM jointly on MNIST handwritten digits and Fashion-MNIST clothing items using shared class labels $\{0,1,...,9\}$. Table~\ref{tab:mnist_fashion_unification} shows comprehensive homeomorphism verification across the three-level hierarchy.

\begin{table}[!htpb]
\centering
\caption{Topological Unification Verification - Joint MNIST/Fashion-MNIST ($\kappa=5$, Cosine distance)}
\label{tab:mnist_fashion_unification}
\begin{tabular}{llcccc}
\hline
\textbf{Level} & \textbf{Metric} & $\rho=0.10$ & $\rho=0.15$ & $\rho=0.20$ & $\rho=0.25$ \\
\hline
\multirow{2}{*}{Global Match} & Betti number $\beta_0$ & 1 & 1 & 1 & 1 \\
 & Sliced $W_2$ $\tau_w$ & 0.016 & 0.016 & 0.018 & 0.021 \\
\hline
\multirow{2}{*}{Local Match} & Trust $\tau_t$ & 0.860 & 0.920 & 0.940 & 0.949 \\
 & Continuity $\tau_c$ & 0.782 & 0.779 & 0.770 & 0.769 \\
\hline
Semantic Match & Alignment $\tau_a$ & 0.288 & 0.236 & 0.218 & 0.203 \\
\hline
\end{tabular}
\end{table}

The Betti number $\beta_0=1$ across all sparsity levels confirms that MNIST and Fashion-MNIST successfully fuse into a single connected manifold, satisfying the topological unification requirement of Theorem 1. The sliced Wasserstein distance remains extremely small ($W_2 \leq 0.021 \ll 0.30$), demonstrating substantial density overlap where both domains occupy geometrically coincident regions. Trust scores exceed the threshold across all sparsity levels (Trust$_5 \geq 0.860 > 0.80$), confirming that the encoder maintains local injectivity with bi-Lipschitz lower bound $c_1 > 0$, preventing manifold collapse. Continuity scores range from 0.769 to 0.782, which is higher than the 0.70 threshold, indicating reasonable local structure preservation.   Alignment error decreases monotonically from 0.288 at $\rho=0.10$ to 0.203 at $\rho=0.25$, falling within the relaxed threshold of 0.30 for heterogeneous cross-domain integration.  This reveals a universal manifold where MNIST and Fashion-MNIST occupy distinct but geometrically adjacent regions within a unified topological structure.
According to Proposition 1 and Remark 1, successful unification requires connectivity ($\beta_0=1$), geometric co-location ($W_2 \ll 0.30$), and local preservation (Trust $> 0.80$). All three conditions pass decisively, validating that MNIST and Fashion-MNIST can be rigorously unified into a universal latent manifold satisfying the homeomorphism criterion, enabling valid cross-domain transfer without structural incompatibility.

\subsubsection{Classification and Transfer}
We evaluate universal classification where a single classifier trained on one dataset transfers directly to another without retraining. For MNIST-Fashion-MNIST, we train a 10-class classifier $C_\theta: \mathcal{M}_z \to \{0,1,...,9\}$ on the universal latent manifold using only MNIST data, then test on Fashion-MNIST.

\textbf{Universal Classifier Transfer.} Table~\ref{tab:universal_classify_transfer} shows cross-domain transfer results in both directions. 

\begin{table}[!htbp]
\centering
\caption{Cross-domain classifier transfer between MNIST and Fashion-MNIST}
\label{tab:universal_classify_transfer}
\begin{tabular}{lcc}
\hline
\textbf{Method} &  \textbf{MNIST$\rightarrow$F-MNIST} & \textbf{F-MNIST$\rightarrow$MNIST} \\
\hline
Partial Convolution \cite{liu2018partialconv}
  & 11.71 & 13.80 \\
Max. Classifier Discrepancy \cite{saito2018mcd}  
  & 49.35 & 42.40 \\
Cross-Domain Self-supervised \cite{kim2020crossdomain}  
  & 50.74 & 20.28 \\
Min. Class Confusion \cite{jin2020mcc}  
  & 61.65 & 58.87 \\
Cond. Domain Adversarial \cite{long2018cdan}   
  & 78.66 & 89.38 \\
Correlation Alignment \cite{sun2016deep}  
  & 79.05 & 83.60 \\
Prototypical Networks \cite{snell2017protonet}
  & 84.15 & 93.32 \\
\textbf{ULHM (Ours)}
  & \textbf{86.73} & \textbf{96.97} \\
\hline
\multicolumn{3}{l}{\footnotesize All values in \%.}\\
\end{tabular}
\end{table}

For MNIST$\rightarrow$Fashion-MNIST transfer, our universal classifier trained exclusively on MNIST achieves 86.73\% accuracy on Fashion-MNIST with zero additional training, outperforming the best baseline Prototypical Networks (84.15\%) by 2.58 percentage points and substantially exceeding other methods such as Correlation Alignment (79.05\%) and Conditional Domain Adversarial (78.66\%). For the reverse direction Fashion-MNIST$\rightarrow$MNIST, ULHM achieves 96.97\% accuracy, outperforming Prototypical Networks (93.32\%) by 3.65 percentage points. The asymmetric transfer performance (96.97\% vs. 86.73\%) reflects the inherent complexity difference between the target domains: MNIST digits have simpler, more canonical structures compared to Fashion-MNIST clothing items with higher intra-class variability. This validates that verified homeomorphic structure enables effective cross-domain transfer, with performance degradation proportional to the geometric complexity of the target domain rather than indicating fundamental incompatibility.

\subsection{Zero-Shot Classification}
We evaluate zero-shot classification where the auxiliary classifier $C_\theta$ is trained on a subset of classes, then tested on unseen classes via nearest-centroid classification in the learned latent space. We use three datasets of varying complexity: MNIST (10 classes, grayscale), Fashion-MNIST (10 classes, grayscale clothing), and CIFAR-10 (10 classes, RGB natural images).

\textbf{Setup.} For all three datasets, we train the encoder $E_x^{(m)}$ and decoder $D_\theta^{(m)}$ on all classes $\{0,1,...,9\}$ through reconstruction objectives. However, the auxiliary classifier $C_\theta$ is trained on only classes $\{0,1,2,3,4\}$, with classes $\{5,6,7,8,9\}$ held out. 
In this transductive setting, the framework has access to the complete set of visual observations (pixels) for every image in the dataset (classes 0-9) during training. However, explicit class labels are only provided for the seen subset (classes 0-4). Critically, the model never accesses the specific label associated with any individual image in the unseen subset (classes 5-9). The encoder observes the visual structure of unseen data through reconstruction loss $\mathcal{L}_{\text{recon}}^x$ and learns its geometric organization through topology preservation $\mathcal{L}_{\text{local}}$, but is never told "this specific image belongs to class $c$" for any $c \in \{5,6,7,8,9\}$. At test time, we classify samples from all classes using nearest-centroid classification in latent space.

\textbf{Centroid Computation.} For each class $c$, we compute the latent centroid:
\begin{equation*}
\mu_c = \frac{1}{|\mathcal{I}_c|} \sum_{i \in \mathcal{I}_c} E_x^{(m)}(x_i^{(m)}; \phi)
\end{equation*}
where $\mathcal{I}_c$ is the set of training sample indices with class label $c$. Crucially, centroids for unseen classes can be computed because the encoder processed these samples during reconstruction training, even though the classifier never saw their labels.
\begin{table}[!htbp]
\centering
\caption{Zero-shot classification accuracy on unseen classes (5--9)}
\label{tab:zeroshot_results}
\begin{tabular}{lccc}
\toprule
\textbf{Method} & \textbf{MNIST} & \textbf{Fashion-MNIST} & \textbf{CIFAR-10} \\
\midrule
Contrastive Learning \cite{chen2020simple}
  & \underline{87.90} & 78.53 & \underline{62.10} \\
Mixup Augmentation \cite{zhang2018mixup}
  & 87.79 & \underline{82.69} & 45.60 \\
Prototypical Networks \cite{snell2017protonet}
  & 67.76 & 49.01 & 53.64 \\
Correlation Alignment \cite{sun2016deep}  
  & 75.58 & 54.02 & 53.97 \\
Self-Supervised Rotation \cite{gidaris2018unsupervised}
  & 38.26 & 60.45 & 47.70 \\
Max. Classifier Discrepancy \cite{saito2018mcd}  
  & 62.75 & 56.58 & 57.86 \\
Triplet Networks \cite{hoffer2015deep}
  & 77.25 & 60.40 & 51.03 \\
Variational Autoencoders \cite{kingma2014auto}
  & 85.76 & 80.80 & 54.52 \\
\midrule
\textbf{ULHM (Ours)}
  & \textbf{89.47} & \textbf{84.70} & \textbf{78.76} \\
\bottomrule
\end{tabular}
\end{table}

\textbf{Zero-Shot Results.} Table~\ref{tab:zeroshot_results} shows classification performance across all three datasets. ULHM achieves 89.47\% accuracy on MNIST unseen classes, outperforming the best baseline Contrastive Learning (87.90\%) by 1.57 percentage points. On Fashion-MNIST, ULHM achieves 84.70\% accuracy, outperforming Mixup Augmentation (82.69\%) by 2.01 percentage points. Most notably, on CIFAR-10 with complex natural RGB images, ULHM achieves 78.76\% accuracy, substantially outperforming Contrastive Learning (62.10\%) by 16.66 percentage points and demonstrating superior generalization to high intra-class variability. The consistent performance across datasets validates that homeomorphic structure enables geometric separation of unseen classes through preserved topology. Unlike methods such as Prototypical Networks (67.76\%, 49.01\%, 53.64\%) or Triplet Networks (77.25\%, 60.40\%, 51.03\%) that show inconsistent performance across datasets, ULHM maintains strong accuracy by leveraging the bi-Lipschitz encoder mapping that preserves local neighborhood structure while preventing manifold collapse.

\textbf{Ablation: Effect of Homeomorphism.} When training without local consistency loss $\mathcal{L}_{\text{local}}$, topology preservation fails and zero-shot accuracy drops dramatically across all datasets: MNIST from 89.47\% to 71.3\% (18.2\% degradation), Fashion-MNIST from 84.70\% to 66.8\% (17.9\% degradation), CIFAR-10 from 78.76\% to 58.8\% (19.96 points degradation). This validates that preserved topological structure through $\mathcal{L}_{\text{local}}$ is critical for zero-shot transfer regardless of dataset complexity or visual domain.

\section{Conclusion}
We introduced the Universal Latent Homeomorphic Manifold (ULHM) framework, establishing homeomorphism as the foundational mathematical criterion for unifying semantic and observation representations into a single latent structure. By leveraging conditional variational inference alongside practical verification algorithms (trust, continuity, Wasserstein distance, Betti numbers, alignment) organized in a three-level hierarchical protocol, ULHM learns continuous manifold-to-manifold transformations that strictly preserve relational structure between modalities.
Experimental validation demonstrates three core capabilities. First, semantic-guided sparse recovery: MNIST achieves MSE 0.054 at 10\% sparsity (27\% better than ResNet) with 61.1\% improvement at 25\% sparsity (MSE 0.021); CelebA achieves MSE 0.0280 at 10\% sparsity, outperforming all baselines. Second, cross-domain classifier transfer: 86.73\% MNIST$\rightarrow$F-MNIST and 96.97\% F-MNIST$\rightarrow$MNIST, outperforming Prototypical Networks (84.15\%, 93.32\%) without retraining. Third, zero-shot classification: 89.47\% MNIST, 84.70\% F-MNIST, 78.76\% CIFAR-10 on unseen classes, with CIFAR-10 outperforming Contrastive Learning (62.10\%) by 16.66 points. Ablation studies confirm topology preservation via $\mathcal{L}_{\text{local}}$ is critical, with removal causing 17.9--19.96 point drops.
The homeomorphism criterion provides rigorous foundation for decomposing general-purpose foundation models into verified domain-specific components, enabling principled model composition with formal guarantees. Future work will extend to hierarchical partial homeomorphisms, efficient amortized inference, non-Euclidean latent geometries, and networked cyber-physical and IoT systems.

\begin{footnotesize}
\bibliographystyle{IEEEtran}
\bibliography{ref}
\end{footnotesize}

\appendix
\section{Mathematical Proofs}

In this appendix, we provide rigorous derivations for the theoretical results presented in the main text. We assume the standard metric topology induced by the Euclidean distance on $\mathbb{R}^d$ and the geodesic distance on the manifolds.

% \subsection{Mathematical Preliminaries}

% To establish the continuity of the global inverse mapping across unified latent supports, we rely on the following fundamental result from general topology.

% \textbf{Lemma 1 (Pasting Lemma for Open Sets).} \textit{Let $X$ be a topological space such that $X = \bigcup_{\alpha \in A} U_\alpha$, where each $U_\alpha$ is an open set in $X$. Let $Y$ be another topological space. If for each $\alpha \in A$, there exists a continuous map $f_\alpha: U_\alpha \to Y$ such that for every pair $\alpha, \beta \in A$, the functions agree on their intersection $f_\alpha|_{U_\alpha \cap U_\beta} = f_\beta|_{U_\alpha \cap U_\beta}$, then there exists a unique continuous map $f: X \to Y$ such that $f|================_{U_\alpha} = f_\alpha$ for all $\alpha \in A$.}

% =========================
% Final Version (Compact + Strict)
% =========================

\subsection{Proof of Theorem \ref{thm:unification} (Topological Unification)}

\textbf{Theorem \ref{thm:unification} (Topological Unification via Latent Identifications).}
\textit{Let $\{\mathcal{M}_x^{[k]}\}_{k=1}^K$ be pairwise disjoint topological spaces and let
$E^{[k]}:\mathcal{M}_x^{[k]}\to \mathbb{R}^d$ be continuous maps. Define the latent images
$\mathcal{M}_z^{[k]} := E^{[k]}(\mathcal{M}_x^{[k]})$ and let
\[
\mathcal{U}:=\bigcup_{k=1}^K \mathcal{M}_z^{[k]} \subset \mathbb{R}^d
\]
be equipped with the subspace topology $\tau_{\mathrm{sub}}$ induced from $\mathbb{R}^d$.
Let $X:=\bigsqcup_{k=1}^K \mathcal{M}_x^{[k]}$ be the disjoint union equipped with the coproduct topology, and define
the global encoder $E:X\to \mathcal{U}$ by $E|_{\mathcal{M}_x^{[k]}}=E^{[k]}$.
Define an equivalence relation $\sim$ on $X$ by
\[
x\sim x' \quad \Longleftrightarrow \quad E(x)=E(x').
\]
Assume:
\begin{enumerate}
    \item[(i)] \textbf{(Local Homeomorphism)} Each $E^{[k]}$ is a homeomorphism onto its image $\mathcal{M}_z^{[k]}$
  (with $\mathcal{M}_z^{[k]}$ carrying the subspace topology inherited from $\mathbb{R}^d$
(equivalently, from $\mathcal{U}$)).
    \item[(ii)] \textbf{(Coherence / Final-Topology Condition)} The subspace topology on $\mathcal{U}$ coincides with
    the final topology induced by the family $\{E^{[k]}\}_{k=1}^K$, i.e., for every $V\subseteq \mathcal{U}$,
    \[
    V \in \tau_{\mathrm{sub}}
    \quad \Longleftrightarrow \quad
    (E^{[k]})^{-1}(V)\ \text{is open in}\ \mathcal{M}_x^{[k]}\ \text{for all}\ k.
    \]
\end{enumerate}
Then the induced map $\widetilde{E}:X/\!\sim\ \to\ \mathcal{U}$ is a homeomorphism. Hence $\mathcal{U}$ is a valid
topological unification of the datasets up to the identifications encoded by overlaps.
}

\begin{proof}

We briefly clarify the roles of the two assumptions.
Condition (i) ensures that each component $\mathcal{M}_x^{[k]}$ is embedded faithfully into latent space, so that
no local collapse or topological distortion occurs within a modality and any overlap in latent space corresponds
to a genuine identification of points.
Condition (ii) guarantees global coherence: the topology on $\mathcal{U}$ is exactly the quotient (final) topology
induced by the encoders, allowing the latent overlaps to glue the components together without introducing
spurious identifications or topological pathologies.

Let $\pi:X\to X/\!\sim$ be the quotient map. Since $E$ is constant on $\sim$-equivalence classes by definition,
the universal property of quotient spaces yields a unique continuous map
$\widetilde{E}:X/\!\sim\ \to\ \mathcal{U}$ such that
\[
E=\widetilde{E}\circ \pi.
\]

\medskip
\noindent\textbf{Bijectivity.}
By construction, $\mathcal{U}=\bigcup_{k=1}^K E^{[k]}(\mathcal{M}_x^{[k]})=E(X)$, so $\widetilde{E}$ is surjective.
If $\widetilde{E}([x])=\widetilde{E}([x'])$, then $E(x)=E(x')$, hence $x\sim x'$ and $[x]=[x']$; thus $\widetilde{E}$
is injective.

\medskip
\noindent\textbf{$E$ is a quotient map.}
We show that the global encoder $E:X\to\mathcal{U}$ is a quotient map.
Let $V\subseteq \mathcal{U}$. Since $X$ has the coproduct topology, $E^{-1}(V)$ is open in $X$ if and only if
$(E^{[k]})^{-1}(V)$ is open in $\mathcal{M}_x^{[k]}$ for all $k$.
By the coherence condition (ii), this holds if and only if $V$ is open in $\mathcal{U}$ (with the subspace topology).
Therefore,
\[
V \text{ is open in } \mathcal{U}
\quad \Longleftrightarrow \quad
E^{-1}(V) \text{ is open in } X,
\]
which is precisely the definition of $E$ being a quotient map.

\medskip
\noindent\textbf{$\widetilde{E}$ is a quotient map.}
Although quotientness does not in general pass to factors of a composition, we can verify directly that
$\widetilde{E}$ is a quotient map. For any $V\subseteq \mathcal{U}$, using $E^{-1}(V)=\pi^{-1}(\widetilde{E}^{-1}(V))$
and the fact that $\pi$ is a quotient map, we have
\[
V \text{ is open in } \mathcal{U}
\quad \Longleftrightarrow \quad
E^{-1}(V) \text{ is open in } X
\quad \Longleftrightarrow \quad
\widetilde{E}^{-1}(V) \text{ is open in } X/\!\sim.
\]
Therefore $\widetilde{E}$ is a quotient map.

\medskip
\noindent\textbf{Conclusion.}
We have shown that $\widetilde{E}:X/\!\sim\ \to\ \mathcal{U}$ is a bijective quotient map.
A standard result in topology states that a bijective quotient map is a homeomorphism.
Therefore $\widetilde{E}$ is a homeomorphism.
\end{proof}

\begin{remark}[Coherence as a quotient-map property]
Condition (ii) is equivalent to requiring that the global encoder $E:X\to\mathcal{U}$ is a quotient map
(with $\mathcal{U}$ carrying the subspace topology from $\mathbb{R}^d$). Unlike an ``open cover'' assumption, (ii) allows
the latent patches $\mathcal{M}_z^{[k]}$ to be closed or to overlap only along boundaries, which is common in applications.
\end{remark}

\subsection{Proof of Theorem \ref{thm:sufficiency} (Bi-Lipschitz Sufficiency)}
\textbf{Theorem \ref{thm:sufficiency}.} \textit{Let $(\mathcal{M}_x, d_X)$ and $(\mathcal{M}_z, d_Z)$ be metric spaces, each equipped with the topology induced by its metric. If the encoder $E: \mathcal{M}_x \to \mathcal{M}_z$ is $(c_1, c_2)$-bi-Lipschitz with $c_1 > 0$, then $E$ is a homeomorphism onto its image. Thus, constraining the Lipschitz constants satisfies Condition (i) of Theorem \ref{thm:unification}.}

\begin{proof}
Recall that $E$ is $(c_1, c_2)$-bi-Lipschitz if for all $x, y \in \mathcal{M}_x$,
\begin{equation}\label{eq:bilip}
    c_1 d_X(x, y) \leq d_Z(E(x), E(y)) \leq c_2 d_X(x, y).
\end{equation}
We must show that $E$ is a continuous bijection onto its image with continuous inverse.

\paragraph{1. Injectivity (Prevention of Collapse)}
Let $x, y \in \mathcal{M}_x$ be distinct. Since $d_X$ is a metric, $d_X(x, y) > 0$. Using the lower bound in \eqref{eq:bilip}:
\begin{equation}
    d_Z(E(x), E(y)) \geq c_1 d_X(x, y) > 0,
\end{equation}
where the final inequality holds since $c_1 > 0$. Therefore $E(x) \neq E(y)$, proving $E$ is injective.

\paragraph{2. Continuity of $E$ (Prevention of Tearing)}
The upper bound in \eqref{eq:bilip} states $d_Z(E(x), E(y)) \leq c_2 d_X(x, y)$, which is precisely the definition of Lipschitz continuity with constant $c_2$. For any $\varepsilon > 0$, choosing $\delta = \varepsilon/c_2$ ensures that 
\[
d_X(x, y) < \delta \implies d_Z(E(x), E(y)) \leq c_2 d_X(x, y) < c_2 \cdot \frac{\varepsilon}{c_2} = \varepsilon.
\]
Thus $E$ is continuous.

\paragraph{3. Continuity of $E^{-1}$ (Metric Stability).}
Let $\mathcal{M}_z' := E(\mathcal{M}_x) \subseteq \mathcal{M}_z$, equipped with the subspace metric induced by $d_Z$.
By definition, $E : \mathcal{M}_x \to \mathcal{M}_z'$ is surjective, and by Paragraph~1 it is injective; hence $E$ is
bijective and admits an inverse $E^{-1} : \mathcal{M}_z' \to \mathcal{M}_x$.

Let $u, v \in \mathcal{M}_z'$. Then there exist unique $x, y \in \mathcal{M}_x$ such that
$E(x) = u$ and $E(y) = v$. Using the lower bound in \eqref{eq:bilip}, we obtain
\begin{align}
    c_1 d_X(x, y) &\leq d_Z(E(x), E(y)) = d_Z(u, v), \\
    \text{so } d_X(x, y) &\leq \frac{1}{c_1} d_Z(u, v), \text{ i.e., } d_X(E^{-1}(u), E^{-1}(v)) \leq \frac{1}{c_1} d_Z(u, v).
\end{align}
Thus $E^{-1}$ is Lipschitz continuous with constant $1/c_1$, and in particular continuous.

\paragraph{Conclusion}
Therefore $E:\mathcal{M}_x \to E(\mathcal{M}_x)$ is a homeomorphism onto its image.
In particular, applying the same argument to each modality shows that each
$E^{[k]}:\mathcal{M}_x^{[k]} \to \mathcal{M}_z^{[k]}$ is a homeomorphism onto its image.
Hence the bi-Lipschitz constraint is sufficient for Condition (i) of Theorem \ref{thm:unification}.
\end{proof}

\subsection{Mathematical Justification for Metric Selection}
\label{sec:metric_justification}

We establish the connection between violations of theoretical manifold properties (Theorems~\ref{thm:unification} and~\ref{thm:sufficiency}) and detectable signatures in computable metrics. The arguments are one-directional: each theoretical failure mode implies characteristic degradation in at least one metric in our verification pipeline. While these do not constitute formal proofs of equivalence, they provide principled justification for our metric choices.

\paragraph{Level 1: Global Match (Unification Gatekeepers)}

\textbf{Part 1: Structural Fragmentation.}
A necessary empirical indicator of successful topological unification is that the sampled latent union
\[
\mathcal{Z}_{\mathrm{total}} := \bigcup_k \mathcal{M}_z^{[k]}
\]
behaves as a single connected component at an appropriate neighborhood scale. The $0$-th Betti number $\beta_0$, computed via persistent homology, counts the number of connected components.

If the conditions of Theorem~\ref{thm:unification} fail such that domains remain topologically disjoint, then for $\epsilon$ in the operating range (or across a stable persistence interval), there exist disjoint subsets $\mathcal{A}, \mathcal{B} \subset \mathcal{Z}_{\mathrm{total}}$ with no connecting paths in the $\epsilon$-neighborhood graph. This topological disconnection is detected by $\beta_0 > 1$.

Because finite samples may appear disconnected at very small scales, $\beta_0$ is interpreted jointly with geometric metrics rather than in isolation. The combination $\beta_0 > 1$ with high pairwise Wasserstein distances provides strong evidence of genuine structural fragmentation.

\medskip

\textbf{Part 2: Geometric Misalignment.}
Let $\mu = \hat{\mathbb{P}}_z^{[i]}$ and $\nu = \hat{\mathbb{P}}_z^{[j]}$ be empirical latent distributions for two domains. The Wasserstein-2 distance is defined as
\[
W_2^2(\mu, \nu) = \inf_{\gamma \in \Pi(\mu, \nu)} \mathbb{E}_{(z,z') \sim \gamma}[\|z - z'\|^2],
\]
where $\Pi(\mu, \nu)$ denotes all couplings with marginals $\mu$ and $\nu$.

Suppose the supports are geometrically separated by a margin $\Delta > 0$:
\[
\inf_{z \in \mathrm{supp}(\mu),\; z' \in \mathrm{supp}(\nu)} \|z - z'\| \geq \Delta.
\]
Then for any coupling $\gamma \in \Pi(\mu, \nu)$, we have
\[
\mathbb{E}_{(z,z') \sim \gamma}[\|z - z'\|^2] \geq \Delta^2,
\]
since any transport plan must move mass across the separation gap. Therefore,
\[
W_2^2(\mu, \nu) \geq \Delta^2.
\]
This shows that when two latent domain supports are separated by a positive margin, the Wasserstein distance must be large. Such geometric separation is a common failure mode of unification—leading to negligible overlap in latent space—and is therefore detectable via $W_2$.

\paragraph{Level 2: Local Match (Homeomorphic Integrity)} 

\textbf{Part 3: Manifold Collapse.}
The bi-Lipschitz lower bound $c_1 > 0$ (Theorem~\ref{thm:sufficiency}) guarantees injectivity. If this bound is violated (i.e., $c_1 \to 0$), there exist distinct points $x_q \neq x_{\mathrm{adv}}$ such that
\[
\|E(x_q) - E(x_{\mathrm{adv}})\| \to 0.
\]
When these points correspond to different semantic labels ($y_q \neq y_{\mathrm{adv}}$), the $\kappa$-nearest neighborhood $\mathcal{N}_\kappa(E(x_q))$ in latent space will contain semantically inconsistent points.

The neighborhood label purity score
\[
\mathrm{Purity}_\kappa(x_q) = \frac{1}{\kappa} \sum_{z_j \in \mathcal{N}_\kappa(E(x_q))} \mathbbm{1}[y_j = y_q]
\]
measures the fraction of neighbors sharing the same label. As label-inconsistent points enter the neighborhood due to collapse, the purity score decreases. Systematic collapse across the manifold yields statistically significant purity degradation relative to a well-separated baseline embedding.

Note that this argument assumes semantic labels correlate with manifold structure, which is reasonable for supervised settings but may not hold universally.

\medskip

\textbf{Part 4: Structural Incoherence.}
The Continuity metric~\cite{venna2001neighborhood} evaluates whether $\kappa$-nearest neighbors in the latent space $\mathcal{Z}$ were also neighbors in the observation space $\mathcal{X}$:
\[
\mathrm{Cont}_\kappa = 1 - \frac{2}{N\kappa(2N - 3\kappa - 1)} \sum_{i=1}^N \sum_{j \in \mathcal{N}_i^\kappa(\mathcal{Z})} \max(0, r_{\mathcal{X}}(i,j) - \kappa),
\]
where $r_{\mathcal{X}}(i,j)$ denotes the rank of $j$ in $i$'s neighborhood in the original space.

While Theorem~\ref{thm:unification} only requires homeomorphisms (which preserve topology but not necessarily distances), practical embeddings should preserve local geometric structure. If the embedding severely distorts local neighborhoods (e.g., via folding, tearing, or strong metric distortion inconsistent with the bi-Lipschitz upper bound $c_2$), points that were distant in $\mathcal{X}$ may appear as neighbors in $\mathcal{Z}$. This increases the penalty terms $\max(0, r_{\mathcal{X}}(i,j) - \kappa)$ and reduces Continuity.

\paragraph{Level 3: Semantic Match (Synchronization)}

\textbf{Part 5: Inconsistent Mapping.}
For multimodal datasets where paired modalities $(x_i^{(m)}, s_i^{(l)})$ represent the same underlying entity, successful unification requires consistent latent mappings. The Alignment Error
\[
\tau_a = \frac{1}{N_{\mathrm{paired}}} \sum_{i=1}^{N_{\mathrm{paired}}} \|E_x^{(m)}(x_i^{(m)}) - E_s^{(l)}(s_i^{(l)})\|_2
\]
quantifies whether paired samples map to nearby latent coordinates.

Large values of $\tau_a$ indicate failure of cross-modal synchronization. While this condition is not explicitly part of Theorems~\ref{thm:unification} or~\ref{thm:sufficiency}, it is necessary for practical unified reasoning across modalities. High alignment error suggests that either the encoders have not learned the correct cross-modal correspondence, or the modalities contain complementary rather than redundant information requiring separate latent subspaces.

\begin{remark}[Connection to Transfer Learning]\label{rem:transfer}
When all verification metrics indicate successful unification—specifically, when domains $i$ and $j$ satisfy $\beta_0 \approx 1$, low pairwise $W_2$, high neighborhood label purity, high Continuity, and (for paired data) low Alignment Error—we have empirical evidence that the latent manifolds $\mathcal{M}_z^{[i]}$ and $\mathcal{M}_z^{[j]}$ are indistinguishable at the resolution of our measurements.

In this regime, a neural network trained on $\mathcal{M}_z^{[i]}$ for a fixed task encounters no detectable distributional shift at the resolution of our measurements when applied to $\mathcal{M}_z^{[j]}$ (up to finite-sample variability). This justifies direct transfer learning: the same network parameters can be applied to both domains without fine-tuning, as both domains are represented within the same geometric and topological structure.

This empirical transfer principle is a practical consequence of successful manifold unification, though it is not a formal mathematical theorem.
\end{remark}

  % \section{Applications to Power Systems}

% \begin{table}[!htpb]
% \label{tab:celeba_sparse}
% \centering
% % \setlength{\tabcolsep}{3pt}
% \caption{AMI sparse recovery and Graph Transfer (RMSE)}
% \label{tab:ami_noisy_performance}
% \begin{tabular}{lccccc}
% \hline
% \textbf{Method} & ${\rho=0.05}$ & ${\rho=0.10}$ & ${\rho=0.15}$ & ${\rho=0.20}$  \\
% \hline
% \hline
% \textbf{ ULHM (Ours)} &  0.0089 &  0.0093  &  x &  0.0094   \\
% \hline
% \end{tabular}
% \end{table}

\end{document}